\documentclass[10pt,journal,compsoc]{IEEEtran}
\usepackage{etoolbox}	%
\newtoggle{vdbeTightFormatting}
\togglefalse{vdbeTightFormatting}
\newtoggle{pnasTemplate}
\togglefalse{pnasTemplate}
\ifCLASSOPTIONcompsoc
  \usepackage[nocompress]{cite}
\else
  \usepackage{cite}
\fi

\ifCLASSINFOpdf
  \usepackage[pdftex]{graphicx}
\else
\fi

\usepackage{amsmath}
\ifCLASSOPTIONcompsoc
  \usepackage[caption=false,font=footnotesize,labelfont=sf,textfont=sf]{subfig}
\else
  \usepackage[caption=false,font=footnotesize]{subfig}
\fi
\usepackage{xspace}
\usepackage{bm}
\usepackage{url}
\usepackage{grffile}	%
\usepackage{textcomp}
\usepackage{color}
\usepackage{colortbl}
\usepackage{booktabs}
\usepackage{subeqnarray}
\usepackage[ruled,vlined]{algorithm2e}
\usepackage{amsfonts}
\usepackage{amssymb}
\usepackage[thmmarks,amsmath,hyperref]{ntheorem}
\usepackage{xfrac}		%
\usepackage{moreverb}	%
\usepackage{framed}		%

\usepackage{pifont}				%
\usepackage[scaled]{beramono}
\usepackage[T1]{fontenc}

\definecolor{shadecolor}{rgb}{0,0,0}%
\definecolor{greyout}{rgb}{0.65,0.7,0.7}%
\definecolor{tbhead}{rgb}{0.97,0.97,0.97}%
\definecolor{a}{rgb}{0.9,0.95,0.95}%
\definecolor{b}{rgb}{0.99,0.99,0.99}%
\definecolor{white}{rgb}{1, 1, 1}%

\def \logicone {{\small\tt 1}\xspace}
\def \logiczero {{\small\tt 0}\xspace}

\newcommand\Warp					{Warp\xspace}

\newcommand{\Hair}{\ifmmode\mskip1mu\else\kern0.08em\fi}
\newcommand\code[1]					{{\bf\ttfamily #1}}

\newcommand\wordLength				{L}

\newcommand{\Z}{\mathbb{Z}}
\newcommand{\abs}[1]{\left|#1\right|}

\theoremstyle{plain}
\theoremsymbol{\ensuremath{\diamondsuit}}
\theoremseparator{.}
\theoremprework{\bigskip\hrule}
\theorempostwork{\hrule\bigskip}

\begin{document}
\title{Probabilistic Value-Deviation-Bounded\\ Source-Dependent Bit-Level\\ Channel Adaptation\\for Approximate Communication}

\author{Bilgesu~Arif~Bilgin~and~Phillip~Stanley-Marbell%
\IEEEcompsocitemizethanks{\IEEEcompsocthanksitem B. Bilgin and P. Stanley-Marbell are with the Department of Engineering, University of Cambridge, Cambridge CB3 0FA, UK.\protect\\
For correspondence e-mail: phillip.stanley-marbell@eng.cam.ac.uk
}%
\thanks{Manuscript received December 2019; revised 11th July 11, 2020.}%
\thanks{Author contributions: PSM formulated the problem of probabilistic value-deviation-bounded source-dependent bit-level channel adaptation, designed the experimental PCB testbed, and assisted with the I2C experiments. BAB developed the methods to calculate the integer value distortion distributions, the search algorithms to find optimal channel adaptation, devised and processed I2C experiment results, derived I2C channel error analysis, and developed the channel adaptation hardware synchronization algorithm. Both authors contributed to writing.}}

\markboth{IEEE Transactions on Computers}%
{Bilgin \MakeLowercase{\textit{et al.}}: Probabilistic Value-Deviation-Bounded Source-Dependent Bit-Level Channel Adaptation for Approximate Communication}
\IEEEtitleabstractindextext{%
\begin{abstract}
Computing systems that can tolerate effects of errors
in their communicated data values can trade this tolerance for
improved resource efficiency. Many important
applications of computing, such as embedded sensor systems, can
tolerate errors that are bounded in their distribution of deviation from
correctness (distortion).  We present a channel adaptation technique
which modulates properties of I/O channels typical in embedded sensor
systems, to provide a tradeoff between I/O power dissipation and
distortion of communicated data.
We provide an efficient-to-compute formulation for the
distribution of integer distortion accounting for the distribution of transmitted
values. Using this formulation we implement our value-deviation-bounded
(VDB) channel adaptation. We experimentally
quantify the achieved reduction in power dissipation
on a hardware prototype integrated with the required programmable channel
modulation circuitry.  We augment these experimental measurements
with an analysis of the distributions of distortions.
We show that our probabilistic VDB channel adaptation can
provide up to a 2$\times$ reduction in I/O power dissipation.
When synthesized for a miniature low-power FPGA intended for use
in sensor interfaces, a register transfer level implementation of the
channel adaptation control logic requires only 106 flip-flops
and 224 4-input LUTs for implementing per-bit channel adaptation on
serialized streams of 8-bit sensor data.
\end{abstract}

}

\maketitle

\IEEEdisplaynontitleabstractindextext
\IEEEpeerreviewmaketitle

\IEEEraisesectionheading{\section{Introduction}\label{sec:introduction}}
\IEEEPARstart{M}{ost} computing systems are designed to prevent errors in computation,
memory, and communication.  Guarding against errors however requires
energy, temporal redundancy, or spatial redundancy and therefore
consumes resources.  But not all systems need to be free of errors:
In some systems, either by explicit design or by the nature of the
problems they solve, system output quality degrades gracefully in
the presence of errors.

Several important applications of computing systems, ranging from
wearable health-monitoring systems to neuromorphic computing
architectures~\cite{merolla2014million} often dissipate a significant
fraction of their energy moving data. For these systems the effects of
errors are best quantified in terms of their integer distance,
rather than using a Hamming distance.  At the same time, the
computations that consume this data can often tolerate errors with
a wide range of integer distances with limited system-level
consequences~\cite{shafique2016cross,
stanleymarbell2018exploiting, 10.1145/2901318.2901347}.

Because electrical communication interfaces do not benefit as much
as processors and other digital logic from semiconductor process
technology scaling, the fraction of system energy dissipated by
sensor data accesses and sensor data movement is currently
large~\cite{hatamkhani2006power, mahony2009future, Malladi:2012}
and will only grow in future embedded sensor-driven and neuromorphic
systems. It is therefore
an important challenge to find ways to reduce the power dissipated
in sensor data acquisition and transfer.  One way to do this is by
exploiting knowledge of the properties and uses of the transported
data, to reduce communication energy expenditure in exchange for,
e.g., bounded inaccuracy of the communicated
values~\cite{stanley2018probabilistic,
Stanley-Marbell:2016:RSI:2897937.2898079, stanley2015efficiency,
kim2017axserbus, stanley2016encoder, huang2015acoco}.

Reducing I/O energy usage in exchange for bounded inaccuracy requires a physical
channel property that provides such a tradeoff. One example of such
a tradeoff is between energy dissipated in the pull-up resistors
on a serial communication interface such as I2C and the error
or erasure rate. The choice of the pull-up resistor value influences
signal integrity at a given I/O speed and at the same time affects
communication power dissipation.

\begin{figure}[t]
\centering
\includegraphics[trim=0cm 0cm 0cm 0cm, clip=true, angle=0, width=0.425\textwidth]{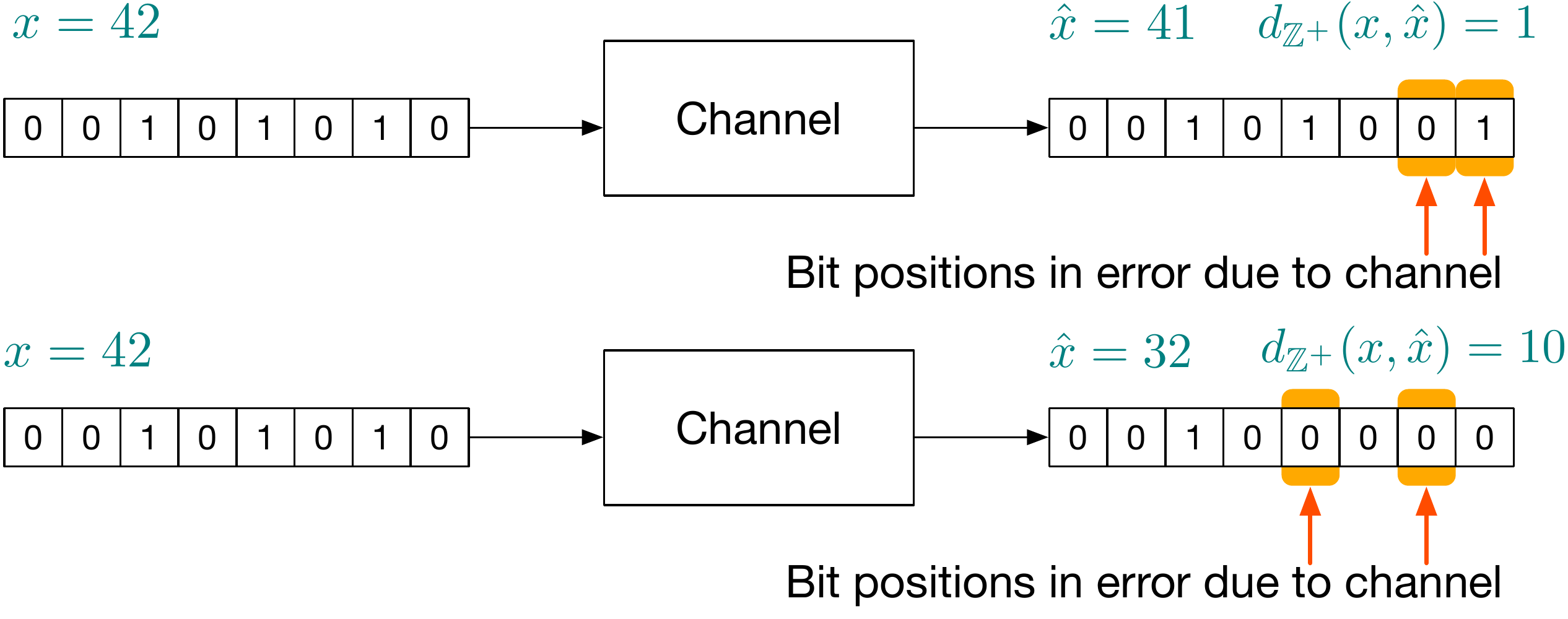}
\caption{Channel model and distance functions: The same number
of bit errors (same Hamming distance) can lead to vastly 
different value deviations (integer distances).}
\label{fig:channel-model}
\end{figure}

\subsection{Contributions}
We exploit value representation and hardware properties of communication
channels to design channel adaptation techniques that
trade lower power dissipation for value-bounded distortion, for a given
bound on its distribution. Given a target bounding distribution, our method
varies the channel bit error rates across the ordinal bit positions in a word to
reduce energy used in transmission in such a way that the distribution of
the resulting distortion is enclosed within the target distribution. We present
a proof-of-concept hardware implementation of the technique which operates
by changing properties of the I2C communication channel for individual bit
positions according to provided channel adaptation specifications. We
calculate these specifications at system design time to generate the appropriate
channel adaptation hardware. The generated channel adaptation hardware is
efficient. Because the technique we present changes
the physical properties of only the channel, its implementation requires no
changes to the transmitter (e.g., a sensor), easing potential adoption.

\textbf{A new technique for approximate communication,
\textit{probabilistic VDB bit-level channel adaptation}
(Section~\ref{section:encoding}):} In a system with integer-representing
binary-valued random variable input $X$ and output $\hat{X}$, let the
random variable $M=d_{\Z+}(X,\hat{X})$ denote the absolute value of
the difference between the integers $X$ and $\hat{X}$.
Figure~\ref{fig:channel-model} shows an example. We refer to the
random variable $M$ as \textit{integer-value distortion}, or just
\textit{distortion}. It takes on instance values $m$ and has tail distribution
$\Pr(M > m)$. We denote integer distance errors tolerated by an application
using an upper bound $\widehat{\mathrm{T}}_{\!M}(m)$ on the
tail distribution of $m$ seen at the output of the system (e.g., the
receiver in Figure~\ref{fig:channel-model}). Given any instance of
the function $\widehat{\mathrm{T}}_{\!M}(m)$, our method
derives an application-specific per-bit channel adaptation 
such that $\Pr(M > m) \le \widehat{\mathrm{T}}_{\!M}(m)$ for all
possible distortion values $m$. In contrast, rate distortion
codes~\cite{Shannon:59, cover2012elements} typically only bound the
expected value of the channel distortion. To facilitate channel adaptation,
we developed a computationally-efficient algorithm for calculating
the distortion distribution of a given channel (i.e., given its bit-error probabilities)
with given input distribution (Section~\ref{subsection:calculation-IVD}).
Using this algorithm, we demonstrate channel adaptation by implementing
the offline search for finding optimal bit-error probabilities with bounded
distortion distribution in the cases when the bit-error probabilities are
identically-distributed over the bit positions and when they are non-identically
distributed (Section~\ref{subsection:demonstration}). The latter
enables bit-level channel adaptation, while the former only achieves
word-level adaptation.

\textbf{A case study for the proof of concept of the proposed method
for the I2C communication channel (Section~\ref{section:I2C}):} We
use pull-up resistance modulation as a means for I2C channel
adaptation and experimentally verify power savings of up to 2$\times$
($0.8\,\mathrm{mW}$ absolute saving) on a custom hardware research
platform (Section~\ref{subsection:I2C-power-dissipation}). This saving in
I/O power is larger than the operation power dissipation of many
state-of-the-art sensors in low-power embedded systems and is
therefore significant. Combining experimental measurements
(Section~\ref{subsection:I2C-power-dissipation} and Section~\ref{subsection:I2C-measurement-results})
with a model of I2C interface circuitry (Section~\ref{subsection:I2C-electrical-parameters}),
we estimate circuit parameters (not always known for commercial integrated circuits)
of our hardware platform (Section~\ref{subsection:I2C-channel-estimation}).
We analytically derive the dependence of the bit-error probabilities
of an I2C channel with given circuit parameter values
on the value of the external pull-up resistance (the channel
modulator). We use this analytic model to calculate the
distortion distribution as a function of pull-up resistance
value and identify the range of resistances for modulation to
have plausible control over bit errors for reliable channel adaptation
(Section~\ref{subsection:I2C-IVD-analytic-derivation}).

\textbf{An analysis of the hardware cost for implementing channel
adaptation (Section~\ref{section:hardware}):} We demonstrate the
feasibility of our probabilistic VDB bit-level channel adaptation
technique by synthesizing the digital logic required to control the
channel adaptation, targeting a small-footprint low-power sensor
interface FPGA. We report the required FPGA resource usage for
sensor sample resolutions common in embedded sensor systems (e.g.,
$8$--$16$ bits).

\subsection{Adaptation versus Source / Channel Coding}
The method we present is hard to classify as a coding scheme in the
conventional nomenclature of the literature. Traditional \textit{source
coding} changes the words transmitted on a channel to adapt
to properties of the inputs (e.g., for compression). Traditional
\textit{channel coding} changes transmitted words to adapt
to properties of the channel (e.g., for forward error correction).
In contrast, the method we present (probabilistic
VDB bit-level channel adaptation) takes the peculiar
approach of changing the properties of the communication channel
for different ordinal positions in a transmitted word in order
to reduce the energy needed to transmit the data. Probabilistic
VDB bit-level channel adaptation does this with the
objective of reducing power dissipation by inducing channel errors
that lead to values at the channel's receiver which are at small
integer distances from those sent by the channel's transmitter.
We present a detailed comparison to the related literature next,
in Section~\ref{section:relatedwork}.

\section{Related Research}
\label{section:relatedwork}
Value-deviation-bounded (VDB) integer codes~\cite{StanleyMarbell:itw09}
were proposed as a means of trading energy efficiency for correctness
in the context of program variable encodings when program variables
can tolerate some distribution of integer value distortions. For
serial communication interfaces such as I2C and SPI, which are
common in energy-constrained embedded systems, serial VDB codes
(VDBS encoding)~\cite{stanley2015efficiency,
Stanley-Marbell:2016:RSI:2897937.2898079} provide a concrete encoding
method.  VDBS encoding is deterministic and has the effect of
re-quantizing the number representation of values. This deterministic
distortion may be undesirable for certain applications: When applied
to images with encoder settings to maximally reduce I/O energy
dissipation, VDBS encoding leads to regular quantization banding
in images. For these and similar applications, encoding methods
that trade energy-efficiency for accuracy and whose induced distortion
is stochastic are therefore desirable.  Existing VDBS encoding
techniques must be implemented at the data transmitter.  Although,
being a re-quantization, they require no decoding, their adoption
requires their integration into data sources such as sensors.
Because the consumers of sensor data, such as microcontrollers, provide
the opportunity for programmability, a variant of VDBS encoding
that could be enforced from the side of consumers of sensor data
could enable widespread adoption of VDBS encoding and its benefits.

Rate distortion theory~\cite{Shannon:59,Shannon:inftheorybook}
investigates the tradeoff of encoding efficiency ({\it rate}) for
deviation of encoded values ({\it distortion}). A {\it distortion
function} or {\it distortion measure} specifies this distortion as
a function of the original data and its encoded form.  The value
deviations we consider in this work are integer distance distortions.
Unlike the convention in rate distortion theory, the method we
present guarantees bounds on the entire distribution of distortions,
rather than only guaranteeing bounds on the expected value of the
distortion.

\textbf{Relation to previous probabilistic VDB methods:}\quad
Prior work~\cite{stanley2018probabilistic} on probabilistic VDB
codes with integer value distortion adopts the same approach of
trying to guarantee bounds on the entire distribution of distortions.
The method we present in Section~\ref{section:encoding} builds on
and improves on that work in several dimensions.
First, we provide a fast method for exact calculation of the distortion
distribution for the general case when bit-error probabilities
may depend on the bits transmitted. This is in contrast with previous~\cite{stanley2018probabilistic}
work, which used an SMT solver to search for optimal bit-level
channel adaptation probabilities. That method only attempts at an approximate
calculation of distortion distribution by considering only up to $k$-bit
concurrent errors for $L$-bit words transmitted. 
Second, we propose an alternative search algorithm  through the space of bit-error
probabilities, which utilizes the exact calculation method we developed.
This method, besides being exact, outperforms the SMT search based
on approximate constraints in terms of computation time.
Third, we verify the feasibility of channel adaptation on I2C systems
by carrying out a proof-of-concept implementation. We demonstrate
power reduction using empirical measurements of power dissipation
and carry out analytical investigation of induced channel errors to explore
the design space. Fourth and finally, we
demonstrate the adoptability of the approach by estimating the
resource usage of the digital logic block required to control the
analog channel adaptation circuitry.

\begin{table}[t]
\caption{Terminology and notation.}
\begin{tabular}{lll}
\toprule
{\bf Notation}			&	{\bf Definition}						&	{\bf Example}\\
\hline
\rowcolor{a}$\wordLength$&	Word length	(in bits)					&	8\\

$x$, $X$			&	Error-free binary value	&	{\scriptsize\tt 0010{\color{greyout}1}0{\color{greyout}1}0}\\

\rowcolor{a}$\hat{x}$, $\hat{X}$	&	Binary value after channel error		&	{\scriptsize\tt 0010{\color{greyout}0}0{\color{greyout}0}0}\\

$k$			&	Number of perturbed bits,	&	2 (grayed-out above)\\
&	Hamming distance $d(x,\hat{x})$	&\\

\rowcolor{a}$u$			&	Conversion from binary	&	$u$({\scriptsize\tt 00101010})$=42$\\
\rowcolor{a}&	 to unsigned integer	&\\

$v$, $V$			&	Error-free integer value, $u(x)$	&	$u$({\scriptsize\tt 00101010})$=42$\\

\rowcolor{a}$w$			&	Integer value after	&	$u$({\scriptsize\tt 00100000})$=32$\\
\rowcolor{a}&	 channel error, $u(\hat{x})$	&\\

$m$, $M$			&	Integer value distortion,	&	10\\
&	$d_{\Z^+}(x, \hat{x}):=\vert v-w\vert$	&\\

\rowcolor{a}$\varepsilon$			&	Signed error vector, $x-\hat{x}$	&	$(0,0,0,0,-1,0,-1,0)$\\

$\mathbb{E}_L$			&	Set of all possible $\varepsilon$ in words	&\\
&	of length $\wordLength$	&\\

\rowcolor{a}$\mathbb{E}_L^m$			&	Set of all possible $\varepsilon$ in words	&\\
\rowcolor{a}&	of length $\wordLength$  with integer value	&\\
\rowcolor{a}&	distortion $m$	&\\

$\mathbb{W}_L$			&	Set of all possible $x$ of length $\wordLength$	&\\

\rowcolor{a}$\mathbb{W}_{L,\varepsilon}$	&	Set of all possible $x\in\mathbb{W}_L$ that	&\\
\rowcolor{a}&	permit a given error vector $\varepsilon$	&\\

$f_X(x)$	&	PMF	of transmitted words $x$	&\\

\rowcolor{a}$f_V(v)$	&	PMF	of transmitted &	See Figure \ref{fig:WISDM_ar_v1.1-user1-fv-distributions}	\\
\rowcolor{a}	&	integer values, $\Pr(V = v)$	&	\\

$\mathrm{F}_{\!V}(v)$	&	CDF of transmitted	&	\\
&	integer values, $\Pr(V \leq v)$	&	\\

\rowcolor{a}$f_M(m)$	&	PMF of integer value	&	See Figure \ref{fig:PMF_and_tailDistribution_IVD}(a)	\\
\rowcolor{a}	&	distortion, $\Pr(M = m)$	&	\\

$\mathrm{F}_{\!M}(m)$	&	CDF of integer value	&	See Figure \ref{fig:PMF_and_tailDistribution_IVD}(b)	\\
&	distortion, $\Pr(M \leq m)$	&	\\

\rowcolor{a}$\widehat{\mathrm{T}}_{\!M}(m)$	&	Constraint tail distribution	&	See Figure~\ref{fig:induced_vs_constraint}\\

$\mbox{T}_{\!M}(m)$	&	Tail distribution, $\Pr(M > m)$	&	$\begin{cases} 1 & m < 4\\ \sfrac{1}{100} & m=4\\ 0 & m \ge 10\end{cases}$\\

\bottomrule
\end{tabular}
\label{table:definitions}
\end{table}

\section{Definitions}
\label{section:definitions}
Table~\ref{table:definitions} summarizes the notation we use in the
rest of this work. 
Let $X$ denote the random variable with instance value $x \in \Z_2^\wordLength$,
an $\wordLength$-bit binary sequence transmitted on a channel.
Let the channel output be $\hat{x} \in \Z_2^\wordLength$ with
corresponding random variable $\hat{X}$.

We consider systems where channel inputs encode
unsigned integers and we define channel errors as the absolute
difference of these integer values. Let $x_i$ denote
the $i$'th bit of $x$ and let $u: \Z_2^\wordLength \rightarrow \Z$
denote the conversion from binary to an integer with
\begin{align*}
u(x)=\sum_{i=0}^{\wordLength-1} x_i 2^i .
\end{align*}
We define the random variable $V:=u(X)$ with instance values $v=u(x)$
and $W:=u(\hat{X})$ with values $w=u(\hat{x})$.
Then, we define the distance $d_{\Z^+}(x, \hat{x})$
between channel input and output as
$$
d_{\Z^+}(x, \hat{x})=\abs{u(x)- u(\hat{x})} = \abs{v- w}.
$$
The random variable $M$ denoting values of
channel distortion, with instance values $m$, takes on values
determined by the distance function $d_{\Z^+}(x, \hat{x})$, i.e., $m=d_{\Z^+}(x, \hat{x})$.
We will use the variable $m$ when we wish to refer to
specific instances of channel integer distortion and we will use
$d_{\Z^+}(x, \hat{x})$ when we want to emphasize the distortion
function.

For each of the random variables defined, we respectively denote
its probability mass function (PMF), cumulative distribution,
and tail distribution by $f$, $\mathrm{F}$, and $\mathrm{T}$ with the
variable as subscript, e.g., $f_M$, $\mathrm{F}_{\!M}$, and $\mathrm{T}_{\!M}$.
In the case of variable $M$, for $\wordLength$-bit words transmitted
over the channel, the domain of $\mathrm{T}_{\!M}$  is $\{0,\ldots,2^\wordLength\!-\!1\}$
and we have
$$
\mathrm{T}_{\!M}(m) = 1- \mathrm{F}_{\!M}(m) =\sum\limits_{i>m}f_M(i), \quad \forall m\in\{0,\ldots,2^\wordLength\!-\!1\}.
$$

Finally, we define the {\it constraint tail distribution}
$\widehat{\mathrm{T}}_{\!M}$ as the distribution of integer
distances $d_{\Z^+}(x, \hat{x})$ which the system
consuming the output of the communication channel can tolerate.
$\widehat{\mathrm{T}}_{\!M}$ has the same domain as $\mathrm{T}_{\!M}$
and it may be any non-increasing function with range $[0,1]$.

\section{Probabilistic VDB Source-Dependent Bit-Level Channel Adaptation}
\label{section:encoding}
We consider a communication channel that transmits integer data represented
by $\wordLength$-bit binary words $x$. Suppose we have a tunable
method at our disposal to induce errors in this communication channel that
lets one vary the error probabilities of each bit in a controllable manner.
We assume that the method of inducing errors on the channel comes with a
benefit (e.g., a reduction in power dissipation) and that we have a
mapping, $\mathrm{B}:\mathbb{R}^{2L}\rightarrow\mathbb{R}$, that quantifies
the benefit for a given set of bit-level error probabilities.

Given the knowledge of bit-error probabilities and the distribution
of input words $x$ to the channel, one can calculate $f_M$, the PMF
for the distribution of induced distortion.
This is equivalent to finding the tail distribution $\mathrm{T}_{\!M}$
of the induced distortion.

Given the foregoing conditions,
probabilistic VDB source-dependent bit-level channel adaptation
consists of finding the correct set of parameters for the mapping $\mathrm{B}$,
i.e., the bit-error probabilities, such that: \ding{202}
$\mathrm{T}_{\!M}(m)\leq\widehat{\mathrm{T}}_{\!M}(m)$ for each
$m\in\{0,\ldots,2^\wordLength \!-\!1\}$ and \ding{203} the benefit $\mathrm{B}$ is maximized.
We first present our method for calculating $f_M$ and then present our search
method for finding bit-error probabilities that satisfy conditions \ding{202} and \ding{203}.

\begin{figure*}[t]
\centering
\subfloat[Probability mass function (PMF) of integer value distortion, $f_M$.]{\includegraphics[trim=2cm 1cm 1.5cm 0cm, clip=true, angle=0, width=0.5\textwidth]{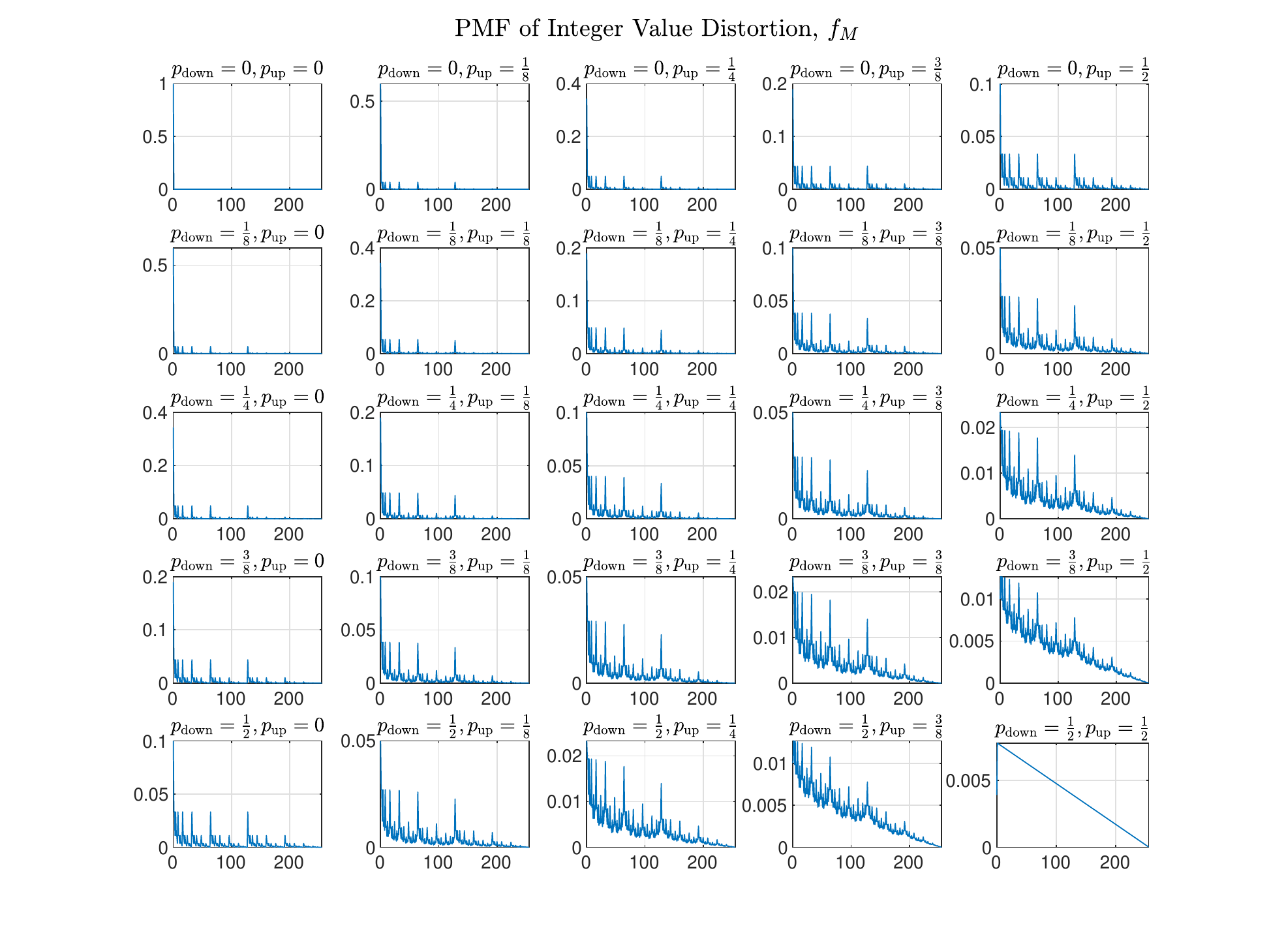}}~~
\subfloat[Tail distribution of integer value distortion, $\mathrm{T}_{\!M}$.]{\includegraphics[trim=2cm 1cm 1.5cm 0cm, clip=true, angle=0, width=0.5\textwidth]{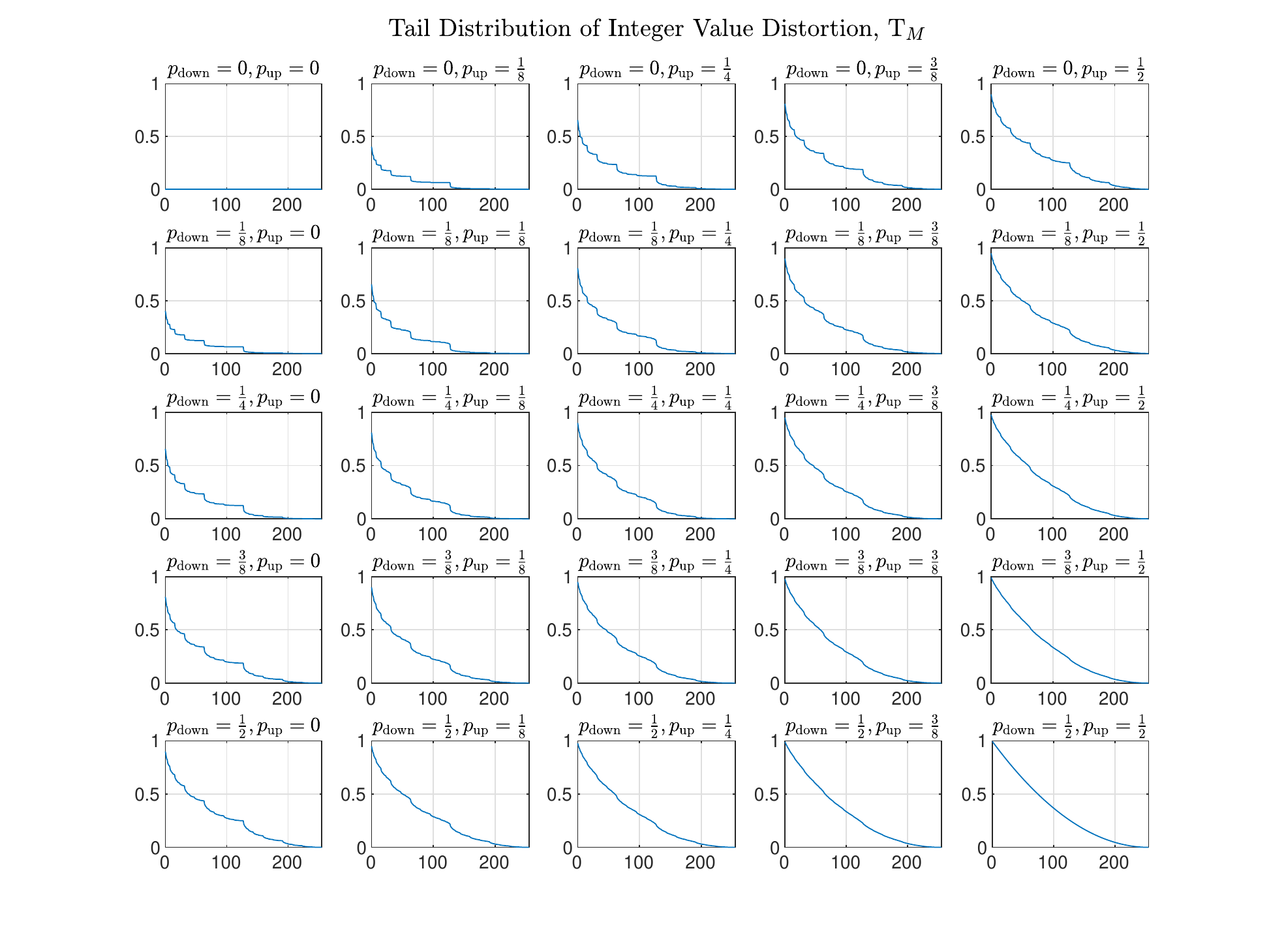}}
\caption{(a) $f_M$ and (b) $\mathrm{T}_{\!M}$ induced by bit-position-independent errors with different probabilities $p_{\mathrm{down}}$ and $p_{\mathrm{up}}$ for the case of 8-bit words and when the input is uniformly distributed. As expected, $\mathrm{T}_{\!M}$ rises as bit-error probabilities increase.}
\label{fig:PMF_and_tailDistribution_IVD}
\end{figure*}

\subsection{Calculating $f_M$}
\label{subsection:calculation-IVD}
Recall, from Table~\ref{table:definitions}, that $f_M$ is the
PMF of integer value distortions that occur during transmission of
$L$-bit words. One can calculate it exactly provided that
one knows the bit-error probabilities suffered during a transmission and
the PMF $f_X$ of input words $x$. Given these,
to calculate $f_M(m)$ for each $m\in\{0,\ldots,2^L\!-\!1\}$, we first characterize all
possible word-level error events resulting in a distortion value of $m$. Then,
for each possible input $x$, we determine the probability of each event
resulting in $m$-distortion given $x$ is the input. The sum of all these
event probabilities is equal to $f_M(m)$.

\noindent\textbf{Characterization of all error events with distortion $m$:}
We can identify all possible word-level error events in $L$-bit words
with the set $\mathbb{E}_{L}$ of all error vectors $\varepsilon$
of length $L$ with entries from the set of error polarities $\{ -1,0,1\}$, i.e.,
\begin{equation*}
\mathbb{E}_{L} := \big\{ \varepsilon \,:\, \varepsilon(i)\in\{ -1,0,1\}, \quad 0\leq i\leq L-1 \big\},
\end{equation*}
where $\varepsilon(i)$ denotes the $i$'th entry of the error vector
$\varepsilon:=x-\hat{x}$. If a word-level error event involves a bit-level
error of \logiczero$\rightarrow$\logicone, \logicone$\rightarrow$\logiczero,
or no error at the $i$'th bit, then the corresponding error vector $\varepsilon$
has $\varepsilon(i)=1$, $\varepsilon(i)=-1$, or $\varepsilon(i)=0$, respectively.
Let the integer value distortion corresponding to a given error vector $\varepsilon$ be
$$
\mathrm{m}(\varepsilon) = \left| \sum\limits_{i=0}^{L-1} \varepsilon(i)2^{L} \right|, \quad \forall \varepsilon \in \mathbb{E}_L.
$$
Given input and output words of the channel $x,\hat{x}\in\mathbb{Z}_2^L$ and
channel error $\varepsilon=x-\hat{x}$, we have $\mathrm{m}(\varepsilon)=d_{\mathbb{Z}^+}(x,\hat{x})$.

Let $\mathbb{E}_{L}^m$ be the set of all possible error vectors
causing $m$-distortion, that is
$$
\mathbb{E}_{L}^m := \left\{ \varepsilon\in \mathbb{E}_{L} \,:\, \mathrm{m}(\varepsilon)=m \right\}.
$$
The set $\mathbb{E}_{L}^m$
characterizes all possible word-level error events with distortion $m$.
Computing $f_M$ requires computing the sets $\mathbb{E}_{L}^m$ for
each $m\in\{0,\ldots,2^L\!-\!1\}$.

The computation of the sets $\mathbb{E}_{L}^m$
makes use of the fact that a word-level error event is a disjoint union
of its constituent bit-level error events (even if distinct bit-level events
may depend on each other). An error at the $i$'th bit introduces
a signed distortion of $2^i$, -$2^i$ or $0$ depending on its polarity.
The distortion resulting from the word-level error event is the absolute
value of the sum of the signed distortions resulting from all constituent
bit-level errors. In analogy to the fact that the distribution of the sum of
two random variables is the convolution of their distributions, one can
construct the sets $\mathbb{E}_{L}^m$ all at once with an $(L\!-\!1)$-step
generalized convolution of set-valued functions.

Let $\alpha$ and $\beta$ be two set-valued functions on $\mathbb{Z}$
such that for each $n\in\mathbb{Z}$ the sets $\alpha(n)$ and $\beta(n)$
contain elements from the same group, with group operation ``$+$''.
The set-valued convolution $f\diamond g$ between
two set-valued functions is again a set-valued function defined by
\begin{equation}\label{generalized_convolution}
\alpha\diamond\beta (n) := \sum\limits_{k+l=n}^{\oplus} \alpha(k)\star \beta(l),
\end{equation}
where $\sum\limits^{\oplus}$ and $\star$ are the summation and
multiplication of sets, respectively. Given two sets
$$\mathbb{A}=\bigcup\limits_{i=1,\ldots,n}\{a_i\} \quad \text{ and } \quad \mathbb{B}=\bigcup\limits_{j=1,\ldots,m}\{b_j\}$$
we define the sets $\mathbb{A} \oplus \mathbb{B}$ and $\mathbb{A} \star \mathbb{B}$ as
$$
\mathbb{A} \oplus \mathbb{B} :=  \mathbb{A} \dot\cup \mathbb{B},
$$
and
$$
\mathbb{A} \star \mathbb{B} := \dot{\bigcup\limits_{\substack{i=1,\ldots ,n \\ j=1,\ldots ,m}}} \{a_i + b_j\},
$$
respectively, where $\dot\cup$ stand for disjoint union.
From the observation that
\begin{equation} \label{set_add_mult_relations}
\left| \mathbb{A} \oplus \mathbb{B} \right| = \left| \mathbb{A} \right| + \left| \mathbb{B} \right| \text{ and } \left| \mathbb{A} \star \mathbb{B} \right| = \left| \mathbb{A} \right| \times \left| \mathbb{B} \right|,
\end{equation}
the set-valued convolution defined by~\eqref{generalized_convolution} reduces to the standard convolution if we only consider the sizes of the sets. For a given set-valued function $f$, if we define $|f|:\mathbb{Z} \rightarrow \mathbb{N}$ by
\begin{equation} \label{integerised_vector-set_fnc}
|f|(m) := \left|f(m)\right|,
\end{equation}
then
\begin{equation} \label{convolution_relation}
\left| f \diamond g \right| = |f| \ast |g|.
\end{equation}

Coming back to the calculation of the sets $\mathbb{E}_{L}^m$,
let the unit positive error vectors $\hat{\varepsilon}_i$
for each $i\in\{0,\ldots, L\!-\!1\}$ be
$$
\hat{\varepsilon}_i(j):=
\begin{cases}
1 &,\text{ when } j=i,\\
0 &,\text{ otherwise},
\end{cases}
\quad0\leq j \leq L\!-\!1,
$$
and consider the vector-set-valued functions $\chi_i$
defined on $\{-2^L\!-\!1,\ldots,0,\ldots,2^L\!-\!1\}$ by
$$
\chi_i(n) =
\begin{cases}
\{-\hat{\varepsilon}_i\}, & \text{ when } n= - 2^i,\\
\{\vec{0}_L\}, & \text{ when } n= 0,\\
\{\hat{\varepsilon}_i\}, & \text{ when } n= 2^i,\\
\emptyset, & \text{ otherwise},
\end{cases}
\quad0\leq i\leq L\!-\!1,
$$
where $\emptyset$ is the empty set and $\vec{0}_L$ is
the zero error vector. For a given $i\in\{0,\ldots, L\!-\!1\}$,
$\chi_i$ encodes the possible error events at the $i$'th bit only
according to their polarities by storing the corresponding
error vector in the set labeled by the signed-integer
distortion the error causes.
Then, the set-valued function, defined by the $(L\!-\!1)$-step
set-valued convolution
\begin{equation} \label{L-1_step_gen_conv}
\mathbb{C}_L := \chi_1 \diamond \cdots \diamond \chi_L,
\end{equation}
encodes all possible word-level error events by storing the
corresponding error vector in the set labeled by its signed-integer
distortion. Consequently, we have
$$
\mathbb{E}_L^m = \mathbb{C}_L(m) \cup \mathbb{C}_L(-m).
$$

\noindent\textbf{Determining event probabilities and finding $f_M$:}
Once we have calculated $\mathbb{E}_L^m$ we can calculate $f_M$
provided that we know the bit-level error probabilities of the channel
and the occurrence probabilities of the input words.
For a given input word $x$, let $p^x_{i,\mathrm{down}}$ and
$p^x_{i,\mathrm{up}}$ denote the probabilities of
\logicone~$\rightarrow$~\logiczero and
\logiczero~$\rightarrow$~\logicone erroneous bit transitions
for the $i$'th bit $x_i$ given it has the correct initial value, i.e.,
given $x_i=1$ for negative polarity error and given $x_i=0$
for positive polarity error.
A given error vector $\varepsilon$ can occur only if the input word
$x$ is in the set
$$
\mathbb{W}_{L,\varepsilon}:= \Bigg\{ x\in\mathbb{W}_{L} :
\begin{cases}
x_i=1&\!\!\!\!\! ,\text{ if } \varepsilon(i)=\!-1,\\
x_i=0&\!\!\!\!\! ,\text{ if } \varepsilon(i)=1,\\
\end{cases}
0\leq i \leq L\!-\!1\Bigg\}.
$$
Hence, the occurrence probability $P_\varepsilon$ of an error
vector $\varepsilon$ is
\begin{equation}\label{eqn:Pw}
P_\varepsilon= \sum\limits_{x\in\mathbb{W}_{L,\varepsilon}}\Pr(x-\hat{x}=\varepsilon | X=x) f_X(x),
\end{equation}
where the probability that $\varepsilon$ occurs given $X=x$ is
\begin{equation}\label{eqn:Pr_wx}
\Pr(x-\hat{x}=\varepsilon | X=x) = \prod\limits_{i=1}^L p^x_i,
\end{equation}
with
\begin{equation}\label{eqn:p_i}
p^x_i=
\begin{cases}
p^x_{i,\mathrm{down}},& \text{ if } \varepsilon(i)=-1,\\
p^x_{i,\mathrm{up}},& \text{ if } \varepsilon(i)=1,\\
1-p^x_{i,\mathrm{down}},& \text{ if } \varepsilon(i)=0 \text{ and } x_i=1,\\
1-p^x_{i,\mathrm{up}},& \text{ if } \varepsilon(i)=0 \text{ and } x_i=0,
\end{cases}
\end{equation}
for $0\leq i \leq L\!-\!1$.

As $\mathbb{E}_L^m$ contains all word-level error events $\varepsilon$
that cause integer value distortion of $m$, summing the error occurrence
probabilities $P_\varepsilon$ over this set gives us the probability that
the integer value distortion is $m$. That is, we have
$$
f_M (m) = \sum\limits_{\varepsilon\in \mathbb{E}_L^m} P_\varepsilon.
$$
Using the PMF $f_M$ we calculate the tail distribution $\mathrm{T}_{\!M}$ as
$$
\mathrm{T}_{\!M} (m) = 1- \sum\limits_{i=0}^m f_M (i).
$$

In the special case of maximal entropy input distribution we have
$$
f_X(x)=2^{-L}, \quad \forall x\in\mathbb{W}_L.
$$
Assuming the individual bit-error probabilities are independent of the transmitted word $x$
(which is not the case for the I2C channel for instance, see Section~\ref{section:I2C}), in view of the
relations~\eqref{eqn:Pr_wx}~and~\eqref{eqn:p_i} and the structure of the set $\mathbb{W}_{L,w}$,
the occurrence probability of $\varepsilon$ given by~\eqref{eqn:Pw} reduces to the product
\begin{equation*}
\begin{split}
P_\varepsilon &= 2^{-L} \!\!\!\! \sum\limits_{x\in\mathbb{W}_{L,\varepsilon}}\Pr(x-\hat{x}=\varepsilon | X=x), \\
&= 2^{-L} \!\!\!\! \prod\limits_{ \varepsilon(i)=-1} \!\! p_{i,\mathrm{down}} \!\! \prod\limits_{ \varepsilon(i)=1} \!\! p_{i,\mathrm{up}} \!\! \prod\limits_{ \varepsilon(i)=0} \!\! (2 - p_{i,\mathrm{down}}-p_{i,\mathrm{up}}).
\end{split}
\end{equation*}

Figure~\ref{fig:PMF_and_tailDistribution_IVD} shows the results of the calculations
carried out in the case of $8$-bit words with maximal entropy input when the error
probabilities are independent of the bit-position, i.e., $p_{i,\mathrm{down/up}}=p_\mathrm{down/up}$
for all $1\leq i\leq 8$.
The plots sweep $p_{\mathrm{down}}$ and $p_{\mathrm{up}}$ in the range of $[0,1/2]$ with resolution $\frac{1}{8}$.
Figure~\ref{fig:PMF_and_tailDistribution_IVD}~(a)~and~(b) respectively depict the dependencies of the
probability mass function and the tail distribution on $p_{\mathrm{down}}$ and $p_{\mathrm{up}}$.
All plots illustrate the expected channel behavior of elevated integer value distortion with increasing bit error probabilities.

\subsection{Searching for Optimal Bit-Error Probabilities}
\label{subsection:demonstration}
In practice, an implemented channel adaptation technique will
only be able to modulate the channel to finitely many distinct states.
This will yield a finite number of possible collections of bit-error probabilities
for all possible words $\{\mathbf{p}^x\}_{x\in \mathbb{W}_\wordLength}$,
where we define
$$
\mathbf{p}^x \!:= (\mathbf{p}^x_\mathrm{down},\mathbf{p}^x_\mathrm{up})\!:=(p^x_{1,\mathrm{down}},\ldots p^x_{L,\mathrm{down}}, p^x_{1,\mathrm{up}},\ldots p^x_{L,\mathrm{up}},).
$$
For each such state, using the corresponding collection of bit-error
probabilities, we can calculate the induced tail distribution
$\mathrm{T}_{\!M} (m)$ as Section~\ref{subsection:calculation-IVD} describes.
Given a benefit functional
$\mathrm{B}:\mathbb{R}^{2\wordLength}\rightarrow \mathbb{R}^+$ we
would then select after an exhaustive search the state
that yields the maximal average benefit
\begin{equation}\label{eqn:average_benefit}
\widetilde{\mathrm{B}} :=\sum\limits_{x\in\mathbb{W}_\wordLength} f_X(x)\mathrm{B}(\mathbf{p}^x),
\end{equation}
while also satisfying the constraint
$\mathrm{T}_{\!M} (m)\leq \widehat{\mathrm{T}}_{\!M}(m)$
for all $m\in\{0,\ldots,2^L\!-\!1\}$.

We demonstrate the results
for the case when the word length is $\wordLength=8$ and assume
that the input words to the channel are uniformly distributed.
We also assume that the bit-error probabilities
are independent of the transmitted word, i.e.,
$p^x_{i,\mathrm{down/up}}=p_{i,\mathrm{down/up}}$ for all
$x\in\mathbb{W}_\wordLength$. In this case, the average benefit
defined by~\eqref{eqn:average_benefit}
reduces to $\widetilde{\mathrm{B}}=\mathrm{B}(\mathbf{p})$.

We consider two sub-scenarios, first when the bit errors are
independent of the bit positions, i.e.,
$p_{i,\mathrm{down/up}}=p_\mathrm{down/up}$ and second when they
depend on the bit positions. We define the benefit functional as
\begin{equation}\label{eqn:benefit_definition}
\mathrm{B}(\mathbf{p}):=\|\mathbf{p}\|^2=\sum\limits_{i=1,\ldots,L} (p^2_{i,\mathrm{down}}+p^2_{i,\mathrm{up}}).
\end{equation}
$\mathrm{B}$ is an increasing function of its argument.
Hence, this choice of $\mathrm{B}$ obeys the principle that the cost (e.g, error suffered) directly
correlates with benefit. \textbf{Searching with bit errors independent of the
bit positions:} In the first scenario we carry out an exhaustive search through the space
of possible bit-error probabilities $(p_\mathrm{down},p_\mathrm{up})\in
[0,\frac{1}{2}]^2$ with a probability resolution of $p_\mathrm{res}=2^{-7}$.
\textbf{Searching with bit errors dependent on the bit positions:} In this scenario,
the space to search through ($\mathbf{p}\in[0,\frac{1}{2}]^{2L}$)
is too large to carry out an exhaustive search with a fine probability resolution
(e.g., $p_\mathrm{res}=2^{-7}$). Thus, we employ a search algorithm with adaptive
probability resolution. Specifically, in the first step, we do an exhaustive search with the initial
probability resolution of  $p^{(1)}_\mathrm{res}\!=\!2^{-2}$ to find an optimal probability vector
$\mathbf{p}_\mathrm{opt}^{(1)}$, that is, one that maximizes the
average benefit $\widetilde{\mathrm{B}}$. The probability
vectors we search through have the form
$$
\mathbf{p}^{(1)}=\mathbf{0}_{2L}+\left((p^{(1)}_\mathrm{res})^{\delta_1},\ldots,(p^{(1)}_\mathrm{res})^{\delta_{2L}}\right),
$$
where $\mathbf{0}_{2L}\in \mathbb{R}^{2L}$ is the $0$-vector and
$\delta_i=0$ or $\delta_i=1$ for $1\leq i\leq 2L$, so that the search involves
$2^{2L}$ probability vectors.
In the subsequent steps, we continually halve the probability resolution
until we reach to the probability resolution of $2^{-7}$
(i.e., $p^{(n)}_\mathrm{res}\!=\!2^{-(n+1)}$, $1\leq n\leq 6$)
and in the $n$'th step, to find $\mathbf{p}_\mathrm{opt}^{(n)}$ that maximizes
$\widetilde{\mathrm{B}}$, we search through the probability vectors of the form
$$
\mathbf{p}^{(n)}=\mathbf{p}_\mathrm{opt}^{(n-1)}+\left((p^{(n-1)}_\mathrm{res})^{\delta_1},\ldots,(p^{(n-1)}_\mathrm{res})^{\delta_{2L}}\right), \, 2\leq n\leq 6.
$$

\begin{figure}[t]
\centering
\includegraphics[trim=7.7cm 2.3cm 8.5cm 2.5cm, clip=true, angle=0, width=0.5\textwidth]{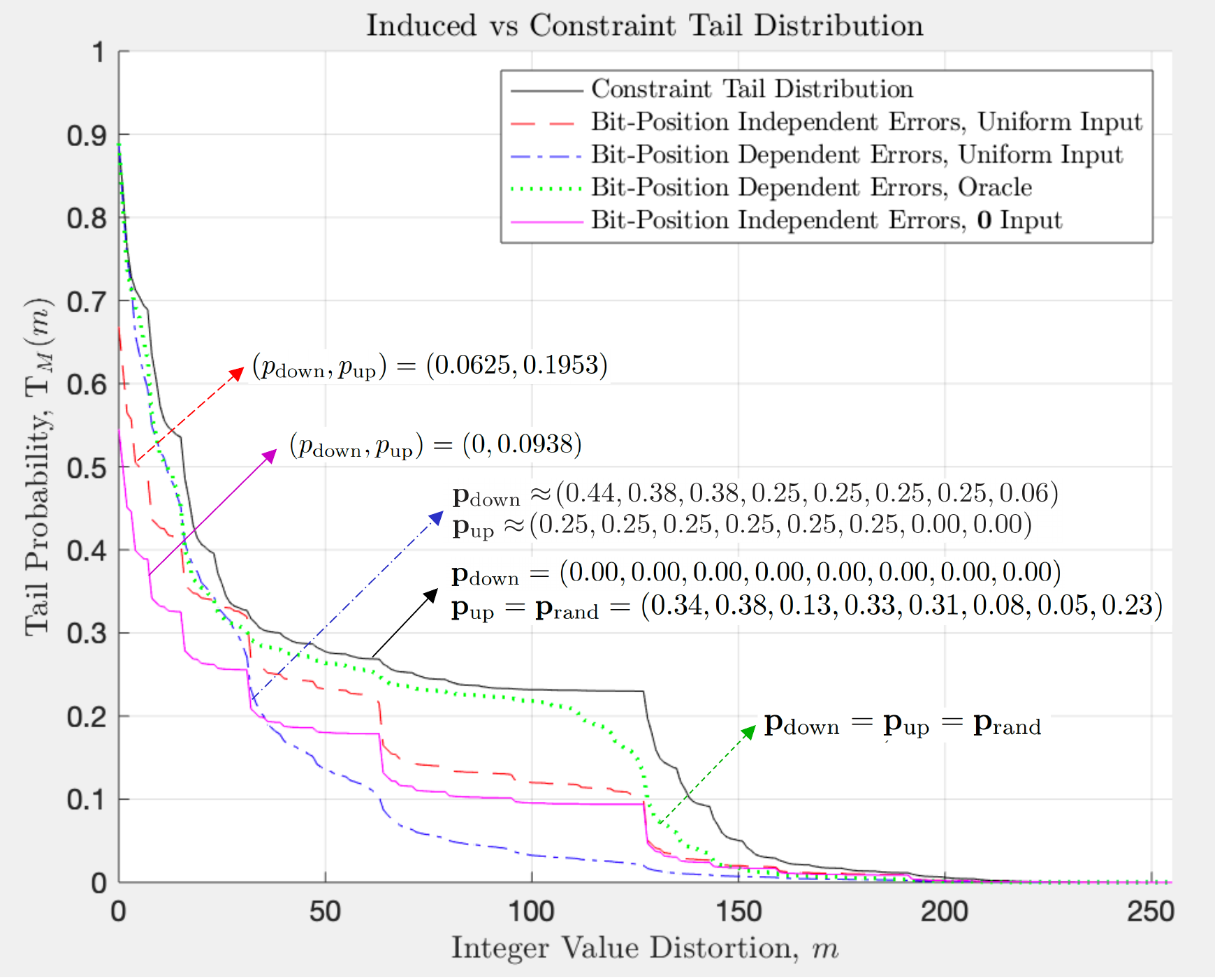}
\caption{Tail distributions of induced integer value distortion corresponding to the bit-error probabilities with maximum average benefit $\widetilde{\mathrm{B}}$ found by the search algorithms in the cases when bit-error manipulation is same for all bit positions with uniform input (red dashed line), distinct at the bit-level with uniform input (blue dash-dot line), in the case of the oracle with uniform input (green dotted line) and when bit-error manipulation is same for all bit positions with $\mathbf{zero}$ input. All induced tail distributions are bounded by the constraint tail distribution $\widehat{\mathrm{T}}_{\!M}(m)$ (black solid line). In the case of the uniform input, bit-level error manipulation achieves more than $3\times$ higher benefit than the bit-independent error manipulation and higher than even what the oracle can achieve.}
\label{fig:induced_vs_constraint}
\end{figure}

\noindent\textbf{Evaluation of search methods:}
To generate an example constraint tail distribution $\widehat{\mathrm{T}}_{\!M}$,
we generate a random probability vector $\mathbf{p}_\mathrm{rand}\in \mathbb{R}^L$
and consider the scenario when only the $\mathbf{0}_L\in\mathbb{R}^L$ is transmitted
over the communication channel. In this scenario the only possible bit-errors
are \logiczero~$\rightarrow$~\logicone errors and we assume that
$\mathbf{p}_\mathrm{rand}$ describes the probabilities of these errors,
i.e., we take $\mathbf{p}_\mathrm{down}:=\mathbf{0}_L$
and $\mathbf{p}_\mathrm{up}:=\mathbf{p}_\mathrm{rand}$. We set the example constraint
tail distribution as the distribution of the integer value distortion suffered in
this scenario. Then, for $0\leq m\leq 2^L\!-\!1$,
the PMF of the suffered integer value distortion is
$$
f_M(m)=\prod\limits_{i=1}^{L} \mathbf{p}_\mathrm{rand}(i)^{u^{-1}(m)_{i-1}} \big(1-\mathbf{p}_\mathrm{rand}(i)\big)^{\big(1-u^{-1}(m)_{i-1}\big)},
$$
where $u$ is defined as in Section~\ref{section:definitions} (see Table~\ref{table:definitions})
and $u^{-1}(m)_i$ corresponds to the value of the $i$'th bit of the binary representation of $m$.
The example constraint tail distribution is given by
$$
\widehat{\mathrm{T}}_{\!M}(m)= 1- \sum\limits_{0\leq i\leq m} f_M(i).
$$

Figure~\ref{fig:induced_vs_constraint} shows for the case when the input is
uniformly distributed the constraint tail distribution together with the induced
tail distributions of integer value distortion corresponding to the benefit-maximizing
bit-error probabilities encountered in our search algorithms.
Figure~\ref{fig:induced_vs_constraint} also plots
the tail distribution of induced integer value distortion when the
bit-error probabilities satisfy
$\mathbf{p}_\mathrm{down}=\mathbf{p}_\mathrm{up}=\mathbf{p}_\mathrm{rand}$.
We refer to this as the \textit{oracle tail distribution}, as it requires
the knowledge of the random vector $\mathbf{p}_\mathrm{rand}$,
which the search algorithms do not possess. The oracle tail distribution
is bounded by the constraint tail distribution, because in the case of error induction with same
bit-error probabilities (for both $\mathbf{p}_\mathrm{down}$ and $\mathbf{p}_\mathrm{up}$)
when input is uniformly distributed as in the case of the constraint tail distribution scenario
(where only $\mathbf{0}_L$ is transmitted and only \logiczero~$\rightarrow$~\logicone bit-errors
can occur) the induced integer value errors by distinct bit-errors may have different signs
and do not compound as in the oracle scenario where distinct bit-errors always have the same sign.
Finally, Figure~\ref{fig:induced_vs_constraint} also plots the tail distribution of integer value distortion
of the exhaustive search for bit position independent errors for the case when the input is always $\mathbf{0}_L$.

In the case of uniform input distribution, the results reveal that the
separate control of error rates in individual bit positions
allows one to induce integer value distortion on the channel
with a larger corresponding average benefit $\widetilde{\mathrm{B}}$ than
achievable by bit-position-independent error induction, as well as than that
the oracle tail distribution achieves. Even though the improvement due to this capability is
undetectable when induced tail distributions are visually compared, in terms of
average benefit maximization the adaptive bit-dependent search algorithm
achieves the largest average benefit with $\widetilde{\mathrm{B}}\approx1.102$ in arbitrary units
compared to $\widetilde{\mathrm{B}}\approx0.336$ of the exhaustive bit-position independent
search (more than $3\times$ improvement) and $\widetilde{\mathrm{B}}\approx1.087$
that the oracle tail distribution achieves. On the other hand, when only $\mathbf{0}_L$ is
transmitted over the channel the bit-position-independent error induction can only achieve
an average benefit of $\widetilde{\mathrm{B}}\approx0.07$ compared to
$\widetilde{\mathrm{B}}\approx0.543$ achievable with bit position dependent error manipulation,
which is the average benefit corresponding to the probabilities of the constraint tail distribution.
In general, independent of the shape of the input distribution and whether the bit errors depend on
the words transmitted, it should be expected that bit-level error manipulation will achieve better
performance in terms of achievable average benefit, because the set of bit-error probabilities admitted
in the case of bit-position-independent error induction is a subset of the set of bit-error probabilities
admitted in the case of bit-position-dependent error induction.

Finally, in comparison to the approximate method of prior work~\cite{stanley2018probabilistic}
for calculating the distribution of integer value distortion for given bit-error probabilities,
the exact calculation method provided in this paper is more computationally efficient.
This fact reflects itself in the computation time the respective search algorithms take
for finding the benefit-maximizing bit-error probabilities inducing a tail distribution of
integer value distortion bounded by a given constraint tail distribution. For the special
case of bit-position-independent bit-error probabilities, while the method of prior work
takes $\sim\!21$ minutes, the method we describe in this section takes only $\sim\!50$
seconds on the same workstation.
\begin{figure*}[t]
\centering
\subfloat[Schematic of an I2C bus with one microprocessor as master and two sensors as slaves. The schematic includes the current and voltage meters we use in our experiments of Section~\ref{subsection:I2C-measurement-results}.]{\includegraphics[trim=0cm 0cm 0cm 0cm, clip=true, angle=0, width=0.75\textwidth]{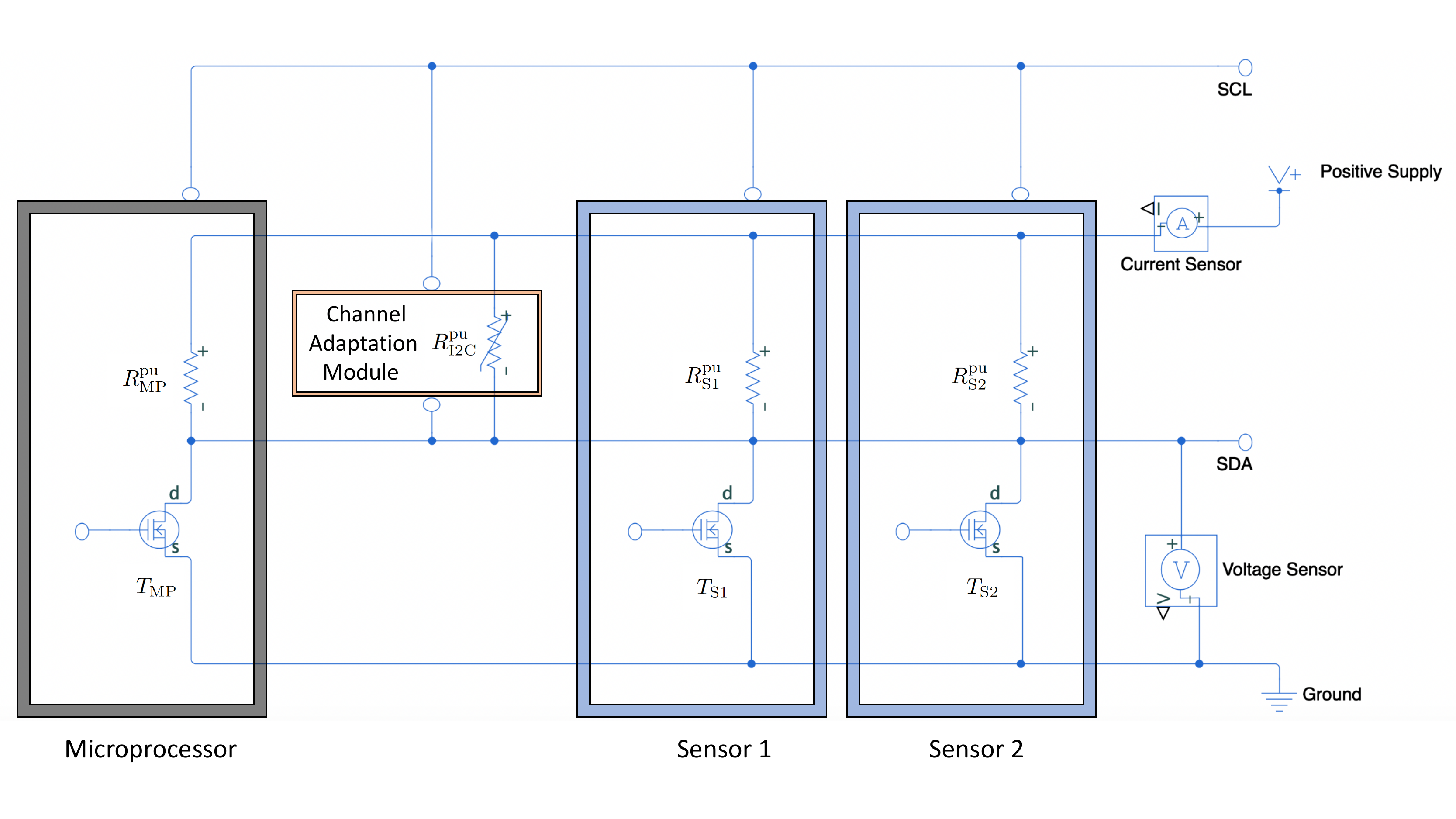}}~~
\subfloat[The equivalent circuit around the I2C data bus during reception event.]{\includegraphics[trim=11.00cm 0.6cm 12.50cm 1cm, clip=true, angle=0, width=0.25\textwidth]{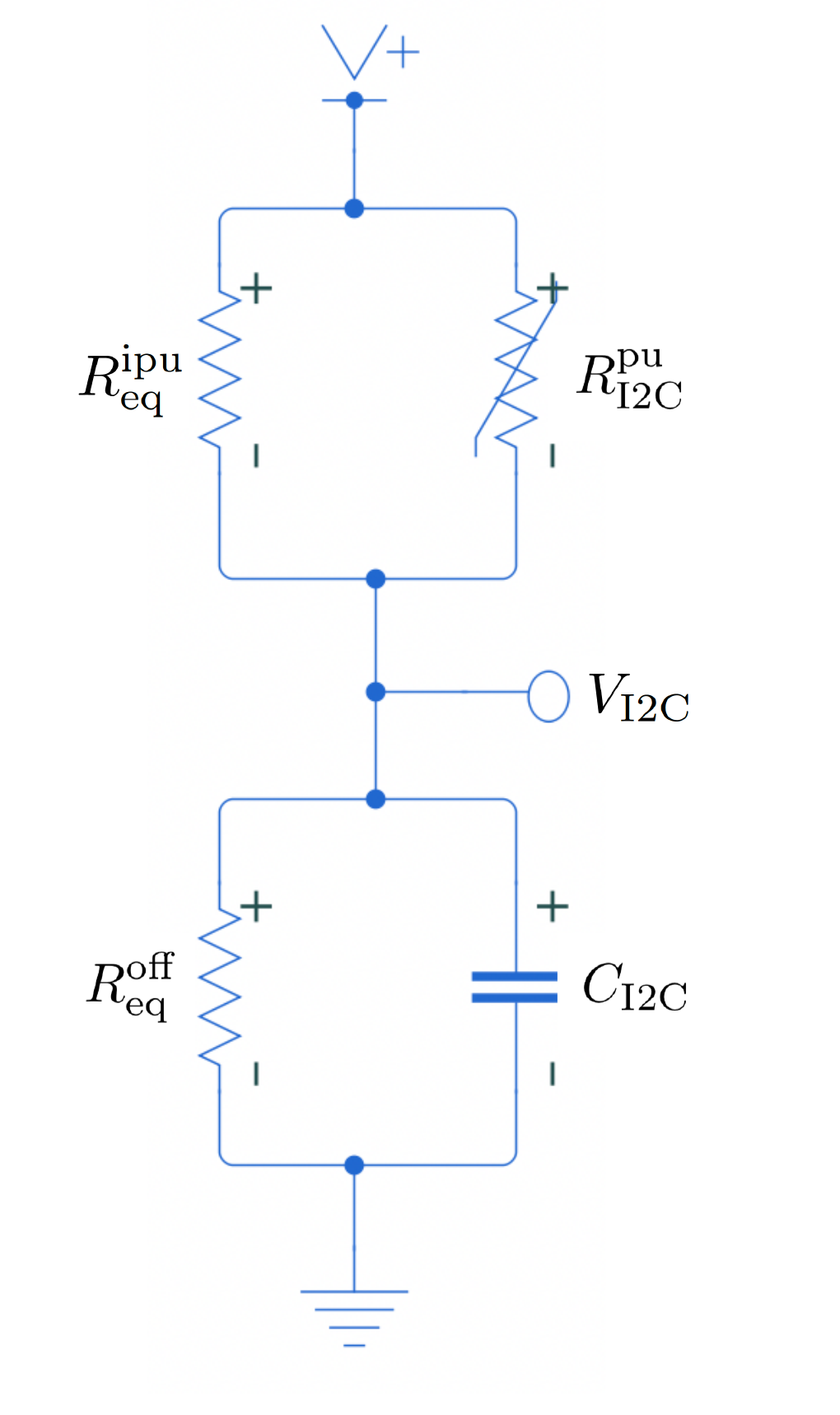}}
\caption{An example I2C system with exposed internal circuitries of agents around the I2C data bus SDA and the experimental setup we used.}
\label{fig:I2C-schematic}
\end{figure*}

\section{Case Study: Channel Adaptation on the I2C Serial Bus}
\label{section:I2C}
The Inter-Integrated Circuit serial bus (I2C) is a serial communication
bus widely used in embedded sensor interfaces and consists of two
signals: a serial clock (SCL) and a serial data line (SDA).
Figure~\ref{fig:I2C-schematic}(a) shows how a processor (the master)
can communicate with multiple sensors (the slaves) over I2C. Data
transfers in I2C occur in multiples of a byte. Drivers on the I2C
bus are open-drain and the I2C standard requires systems implementing
it to provide pull-up resistors on both SDA and SCL. To implement
the channel adaptation method of Section~\ref{section:encoding},
we implement these pull-up resistors with two digitally-controlled
potentiometers (DCPs). Because of the pull-up resistors, when the I2C
bus is idle, both SDA and SCL voltages are close to the I/O supply
voltage. To pull SDA or SCL low, either the master or the slave sinks
the current through its drive transistor to ground.

\subsection{I2C Electrical Parameters}
\label{subsection:I2C-electrical-parameters}
Typically, designers choose a significantly larger value for the pull-up
resistance $R_\mathrm{I2C}^\mathrm{pu}$ compared to the on-resistance
$R^\mathrm{on}_T$ of the transistor in order to achieve logic-\logiczero
voltage close to zero.
Furthermore, for the bus to be able to charge/discharge to these
voltage levels within an I2C clock cycle $t_\mathrm{I2C}$ one
requires the charging/discharging time constants of the I2C data
bus $\tau_\mathrm{up}$ and $\tau_\mathrm{down}$ to be comparably
small. As transistors with both very small on- and very large off-resistances
are available, choosing $R_\mathrm{I2C}^\mathrm{pu}$
small enough to be able to charge the
bus in time is the main constraint determining its value from the
perspective of reliable transmission.  On the other hand, the main
source of power dissipation in I2C data transmission is due to
current flow through $R_\mathrm{I2C}^\mathrm{pu}$ during a logic~\logiczero
transmission. Consequently, during transmission,
using values of $R_\mathrm{I2C}^\mathrm{pu}$ higher than those
used for reliable operation will reduce reliability, but it will also reduce
power dissipation.

Reduction in transmission reliability corresponds to an increase in
bit-error probabilities, whereas reduction in power dissipation
corresponds to an increase in power savings. Accordingly, the benefit
functional $\mathrm{B}$ defined in Section~\ref{section:encoding},
which takes the bit-error probabilities as input and whose value
translates to power savings within the I2C channel adaptation framework,
should be an increasing function of its arguments.
The definition~\eqref{eqn:benefit_definition} in Section~\ref{subsection:demonstration}
obeys this principle.

\subsection{Implementing Bit-Level Channel Adaptation for I2C}
We implement source-dependent bit-level channel adaptation for the
I2C channel, establishing a tradeoff between transmission reliability
and power dissipation by manipulating the value of the pull-up
resistance $R_\mathrm{I2C}^\mathrm{pu}$ during each bit transmission
of an I2C transaction. Here, we focus on the analog electrical behavior
of the bus for different per-bit values of the pull-up resistance achieved
by the combination of digital modulator logic and the variable resistor.
We present the digital logic for controlling the DCP in Section~\ref{section:hardware}.

The channel adaptation module in Figure~\ref{fig:I2C-schematic}(a)
takes as input the I2C SCL and SDA lines and modulates the value
of the pull-up resistor on a per-bit-clock basis. For each sensor
$S$ on the I2C channel that transmits data in chunks of $\wordLength_S$
bits, the channel adaptation module contains offline-calculated
pull-up resistance selections corresponding to resistances
$\{R_\mathrm{I2C}^{\mathrm{pu},i}\}_{i=1}^{\wordLength_S}$, which
will lead to associated bit error probabilities that are permissible
for that bit position (calculated in Section~\ref{section:encoding}).
The channel adaptation module synchronizes with the I2C transaction
and switches the pull-up resistance to $R_\mathrm{I2C}^{\mathrm{pu},i}$
at the beginning of transmission of the $i$'th bit of each data chunk of size
$\wordLength_S$, $1\leq i\leq\wordLength_S$.
The switching of the pull-up resistances on a per-bit-clock basis
requires the channel adaptation module to switch the DCP at a
frequency much higher than the clock frequency of the
I2C transaction, which may go up to $5\,$MHz (referred to as the ultra-fast mode).
On the other hand, the CMOS switches on a typical application-specific integrated circuit
with a $90\,$nm process can switch their output in $120\,$ps and can do so every
$10\,$ns, which permits bit-level channel adaptation even for ultra-fast mode
I2C transactions with the pull-up resistance switching occupying less than
$0.1\%$ of an I2C clock cycle.

\begin{figure}[t]
\centering
\includegraphics[trim=0cm 0cm 0cm 0cm, clip=true, angle=0, width=0.16\textwidth]{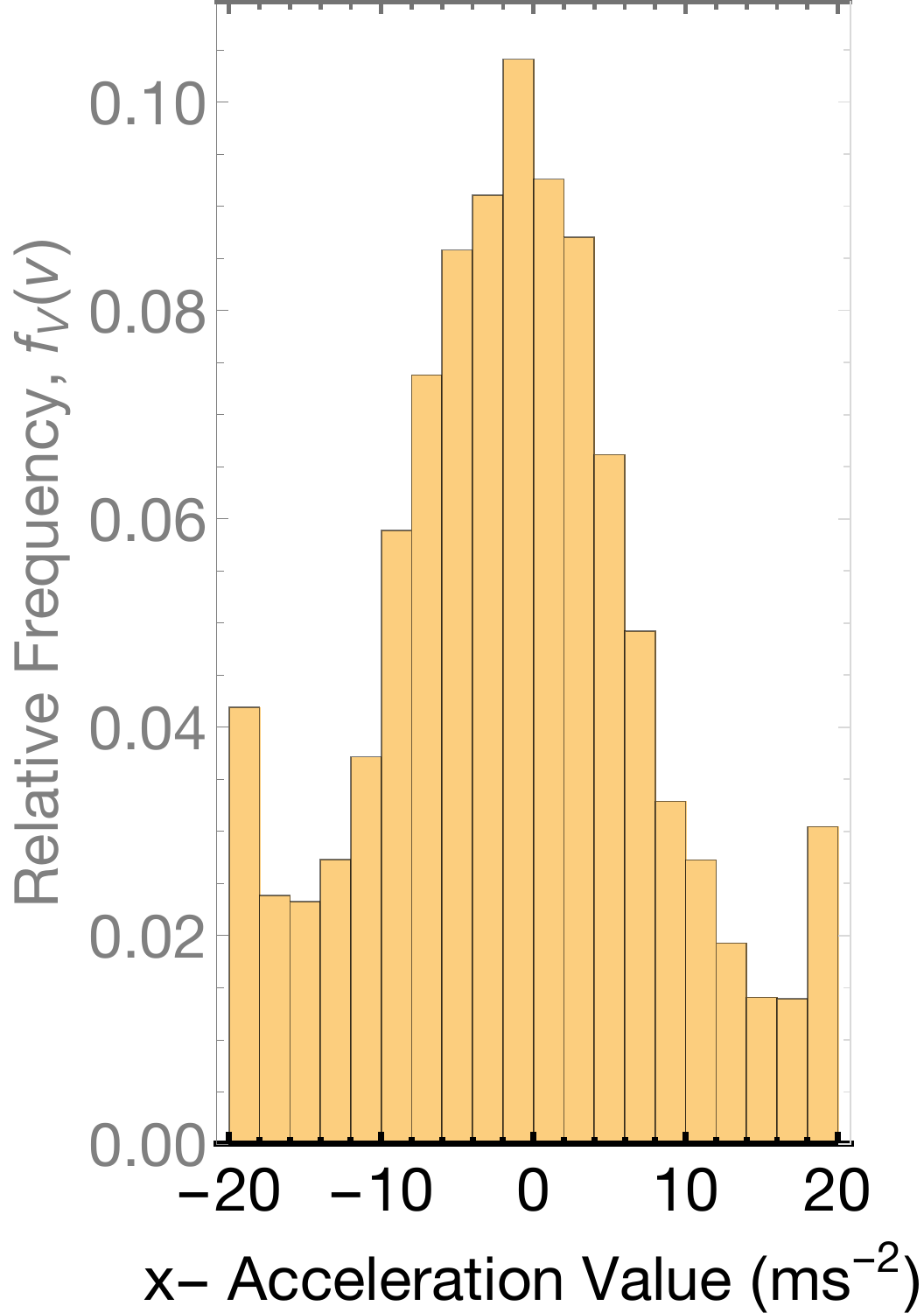}~
\includegraphics[trim=0cm 0cm 0cm 0cm, clip=true, angle=0, width=0.16\textwidth]{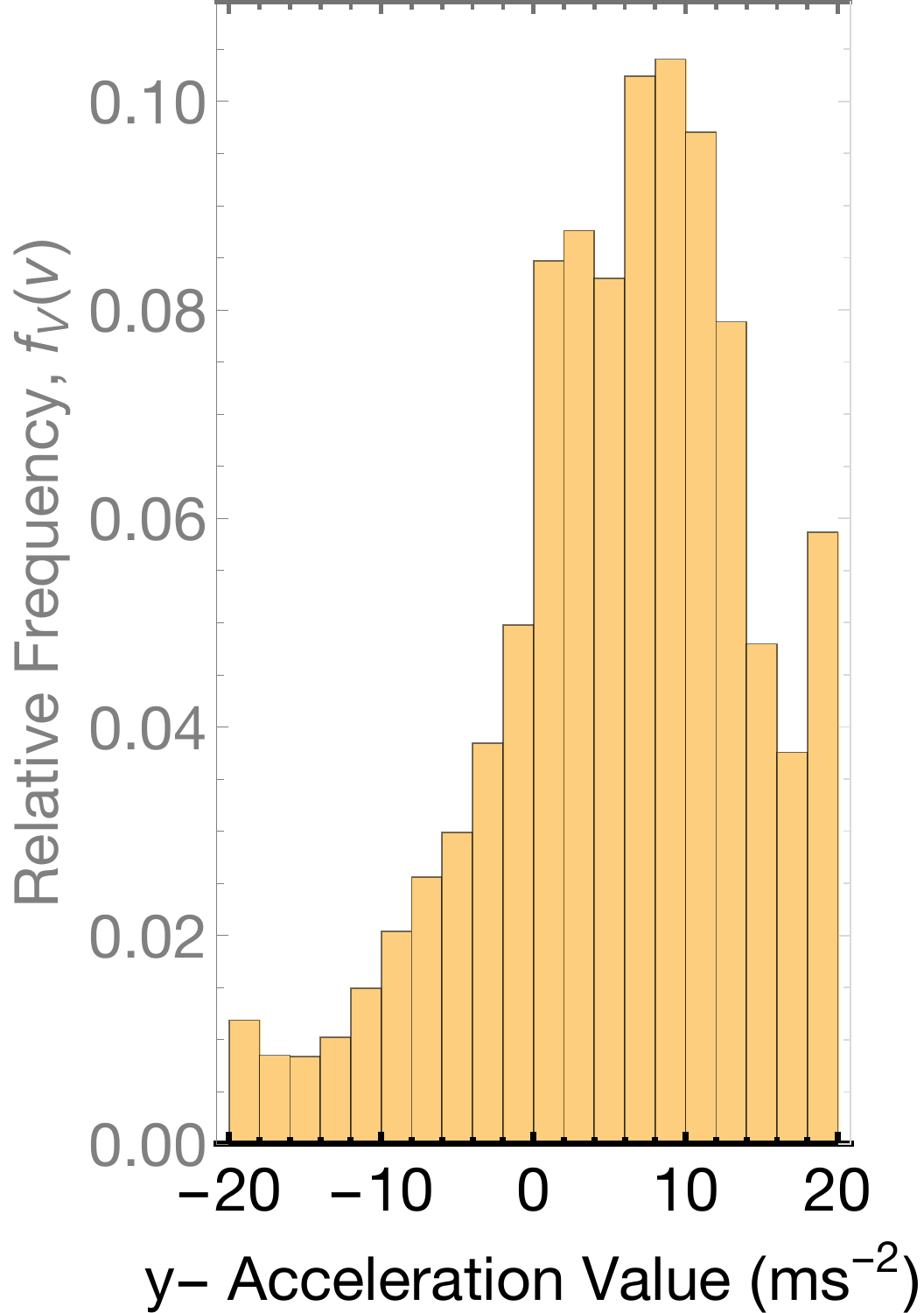}~
\includegraphics[trim=0cm 0cm 0cm 0cm, clip=true, angle=0, width=0.16\textwidth]{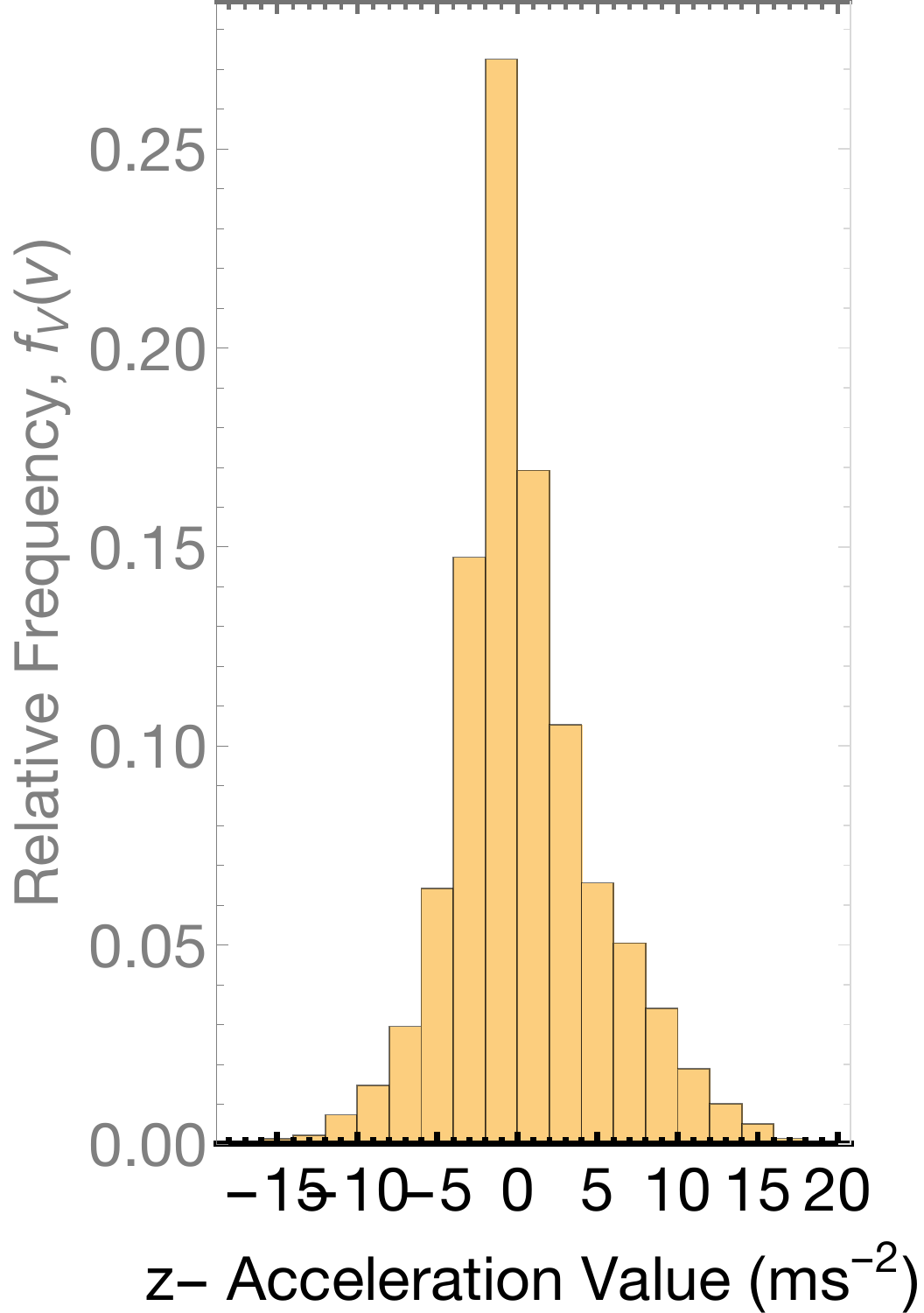}
\caption{Empirical probability mass functions (i.e., $f_{\!V}(v)$
see Table~\ref{table:definitions}) for 29\,978 samples (25 minutes of
sensor data sampled at 20\,Hz) from the three axes of an accelerometer
worn by a person engaged in a variety of
activities~\protect\cite{Kwapisz:2011}.}
\label{fig:WISDM_ar_v1.1-user1-fv-distributions}
\end{figure}

The user input of our method is the expected sensor data distribution
$f_V$ and an upper bound tail distribution $\widehat{\mathrm{T}}_{\!M}$
on the permitted integer value distortion.
Figure~\ref{fig:WISDM_ar_v1.1-user1-fv-distributions} shows an
example of real-world distributions of the error-free integer values
$u(x)$ (see Section~\ref{section:definitions}). Using these inputs
in an offline step performed at system design time, we derive a
specification of $\wordLength_S$ pull-up resistance selections.
The resulting communication on the I2C bit-level-adapted channel
which switches pull-up resistance according to this specification yields
a tail distribution of integer value distortion $\mathrm{T}_{\!M}$
globally bounded by $\widehat{\mathrm{T}}_{\!M}$. The resulting
approximate channel consumes less power than the unadapted channel.

To be able to find optimal pull-up resistance values for I2C channel
adaptation given the input distribution and user-defined bound on
distortion distribution and to estimate the resulting power savings,
we need to characterize the dependence of power dissipation and
induced distortion distribution on pull-up resistance value. We 
measure the power dissipation dependence experimentally. This
also lets us estimate the parameters of the equivalent circuit around
the I2C data bus during reception of logic~\logicone provided in
Figure~\ref{fig:I2C-schematic}~(b). For the dependence of integer
value distortion incurred during transmission we present an analytic
model in Section~\ref{subsection:I2C-IVD-analytic-derivation}.

\begin{figure}[t]
\centering
{\includegraphics[trim=1cm 0cm 4cm 0cm, clip=true, angle=0, width=0.5\textwidth]{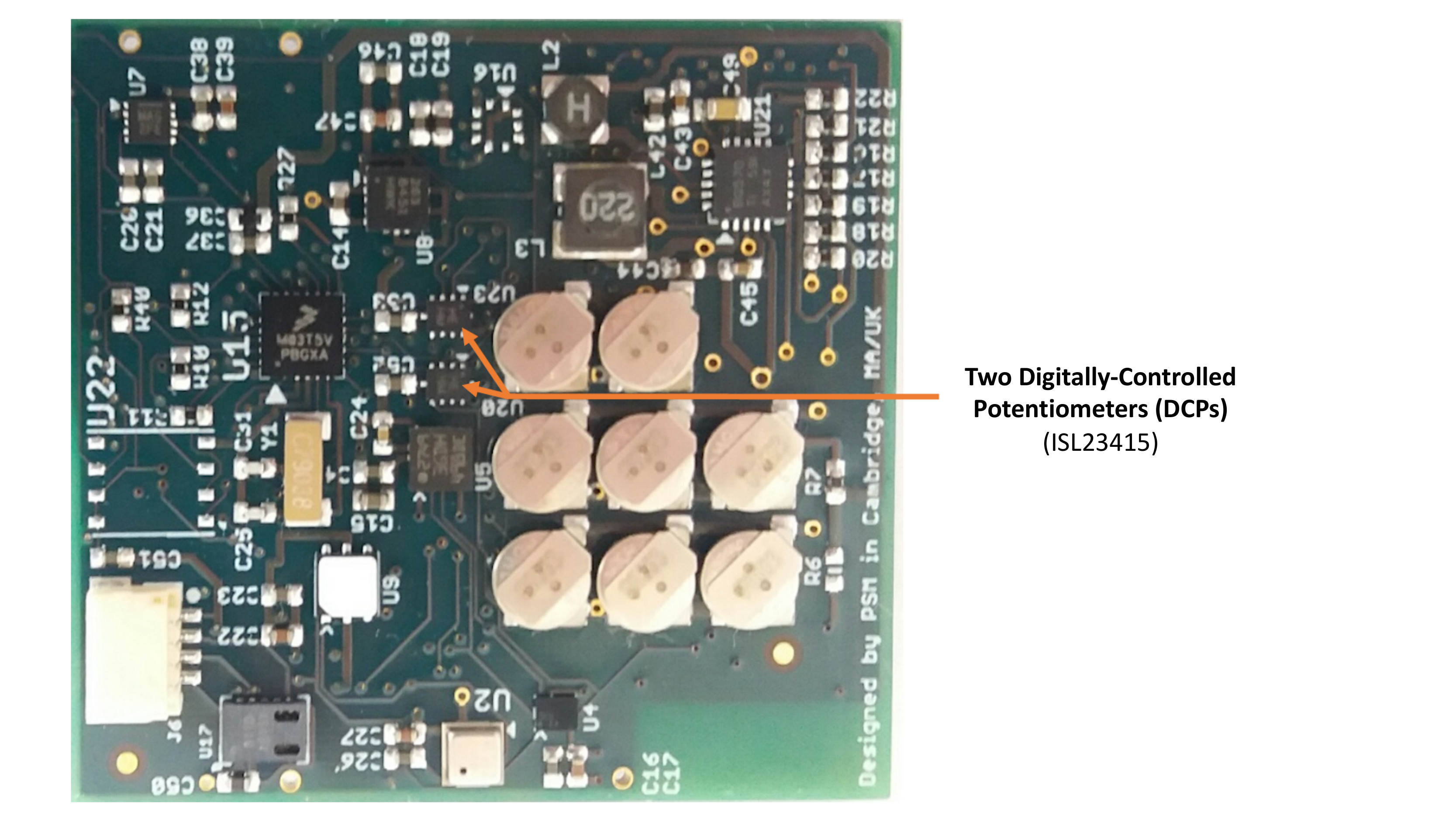}}
\caption{The \Warp embedded hardware-research platform contains two DCPs configurable to serve as pull-up resistors to its I2C channel.}
\label{fig:termination-in-warp}
\end{figure}

\begin{figure*}
\centering
\subfloat[Raw current data for all the bytes transmitted for the experiment at two extreme values of the pull-up resistance.]{\includegraphics[trim=0cm 0cm 0cm 0cm, clip=true, angle=0, width=0.33\textwidth]{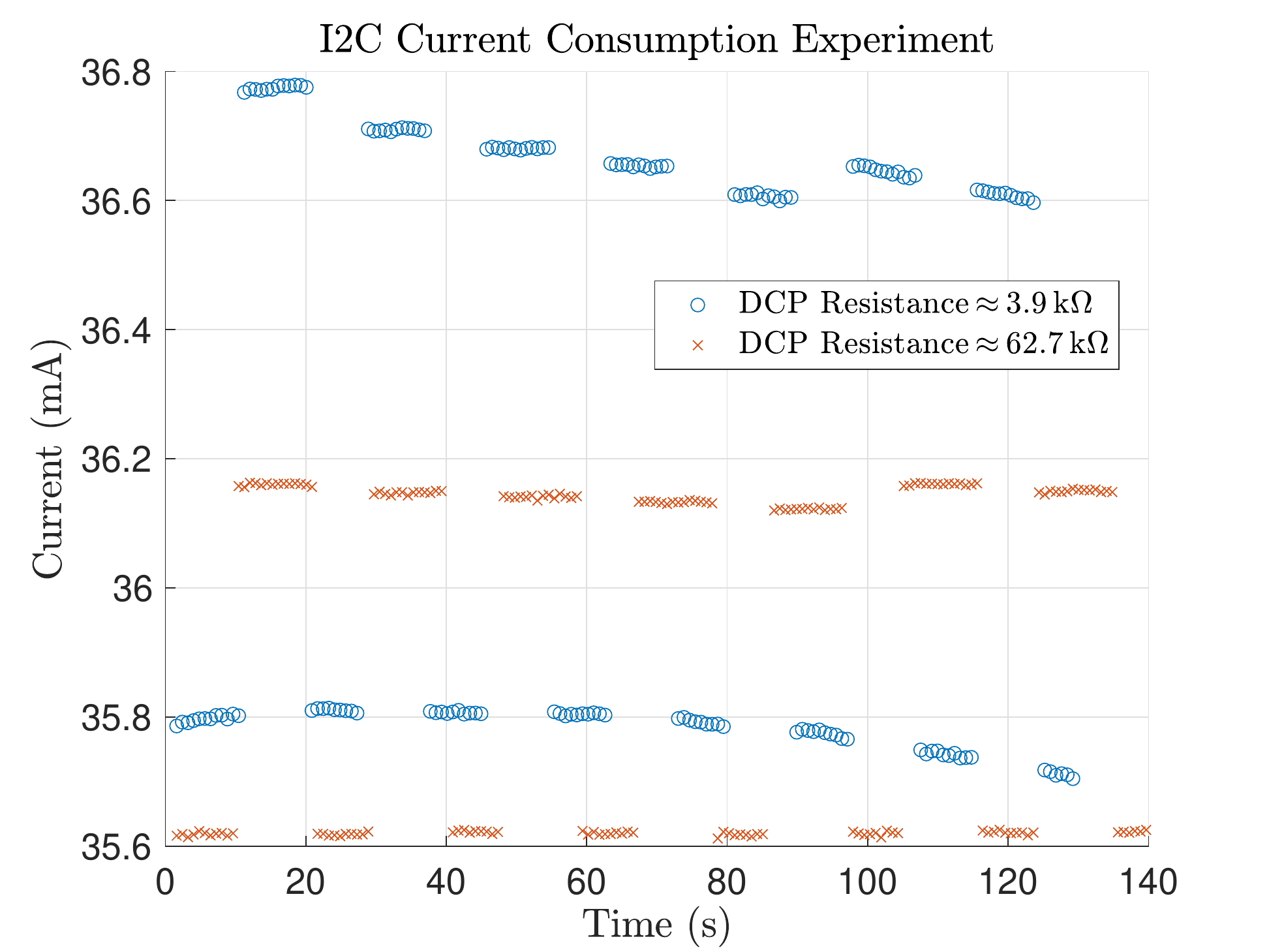}}~~
\subfloat[Average I2C power consumption as a function of pull-up resistance for transmitted bytes with varying duty cycles.]{\includegraphics[trim=0cm 0cm 0cm 0cm, clip=true, angle=0, width=0.33\textwidth]{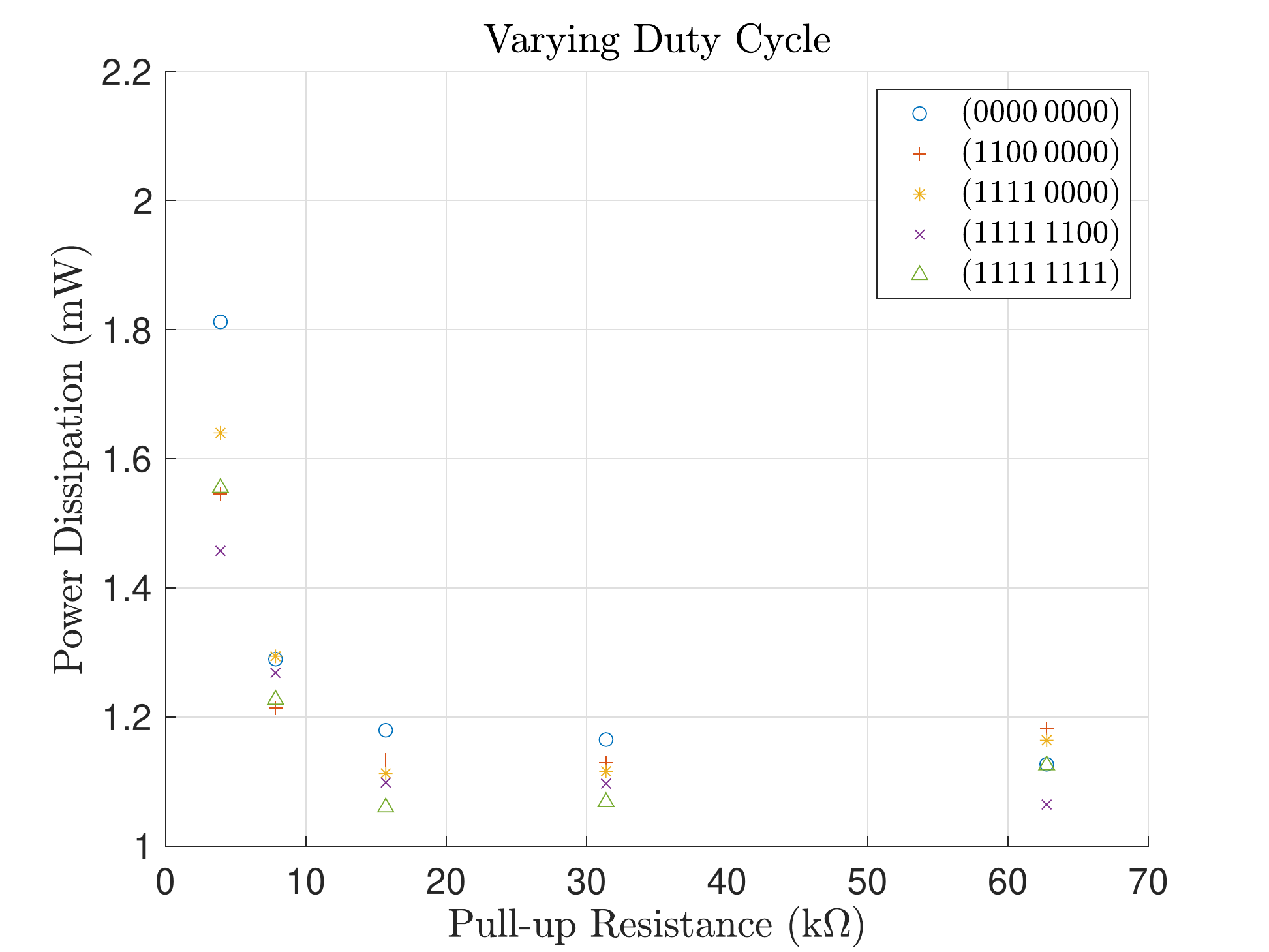}}~~
\subfloat[Average I2C power consumption as a function of pull-up resistance for transmitted bytes with varying logic switch frequency.]{\includegraphics[trim=0cm 0cm 0cm 0cm, clip=true, angle=0, width=0.33\textwidth]{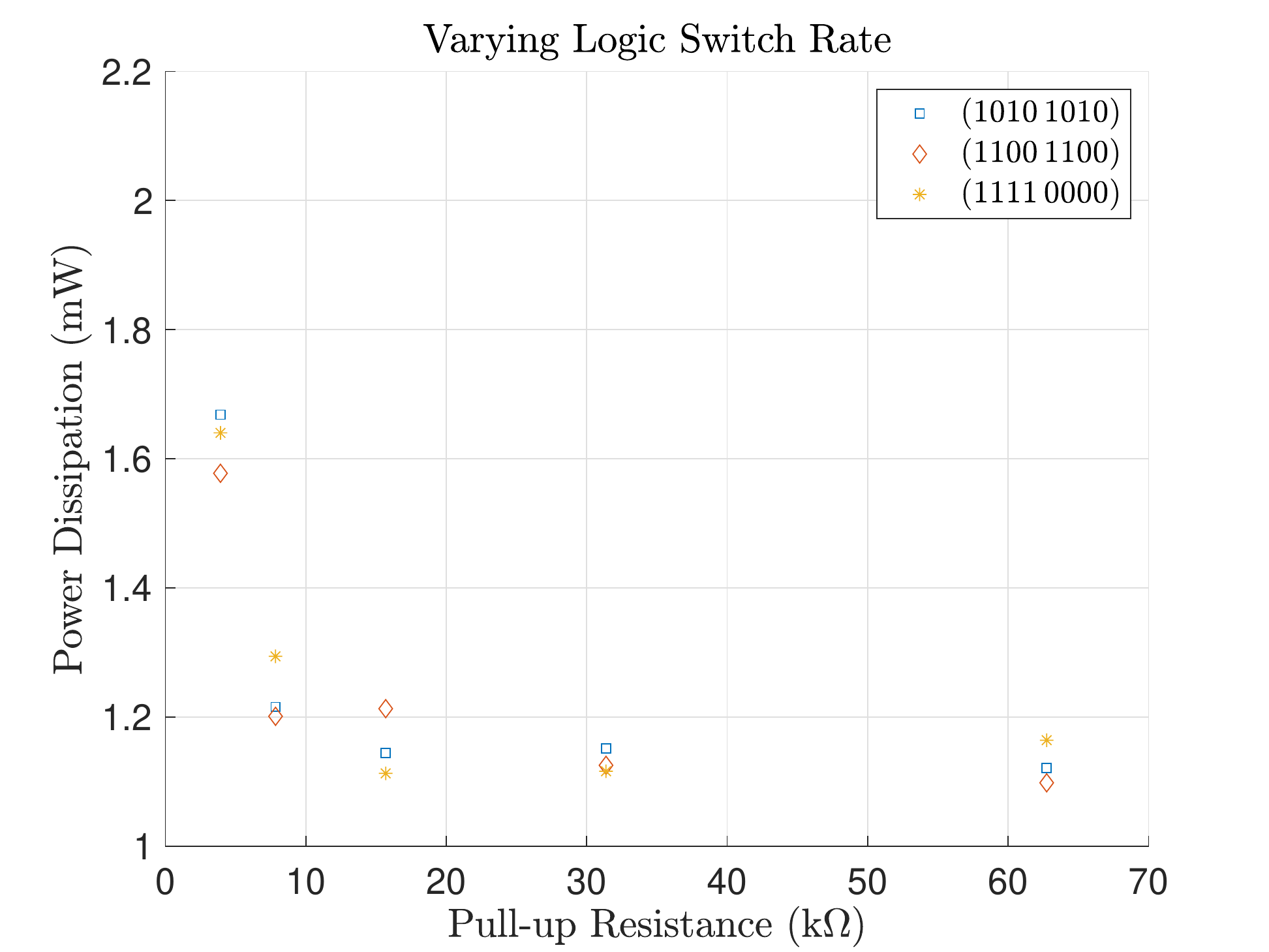}}
\caption{Power measurement experiment results show clear inverse dependence of I2C power consumption on the I2C pull-up resistance value.}
\label{fig:I2C-experiment}
\end{figure*}

\subsection{Setup for Measuring the I2C Power Dissipation Dependence on Pull-up Resistance}
\label{subsection:I2C-power-dissipation}
To determine the dependence of I2C bus power dissipation on the value
of the pull-up resistor, we ran experiments on the I2C communication
interface in \Warp~\cite{8959350}, an open-source hardware platform
for research into tradeoffs between accuracy and power
dissipation (see Figure~\ref{fig:termination-in-warp}).

We use two ISL23415 DCPs~\cite{ISL23415_datasheet} as
configurable pull-up resistors on the SDA and SCL lines of the I2C bus.
The ISL23415 is an integrated circuit that occupies an area of less than
$2\,\mathrm{mm}^2$. It dissipates less than $0.5\,\mu\mathrm{W}$
of power over its CMOS switches when switching between different resistance
settings. Because we cannot control the ISL23415 DCPs to switch
resistance values between I2C clock cycles, we use them primarily
to demonstrate the feasibility and potential power savings for the case
where we use the same pull-up resistance value for all clock cycles.
A custom integrated circuit combining the digital interface from
Section~\ref{section:hardware} with the internals of a device such
as the ISL23415 would be primarily an engineering task and would
not provide much additional research insight.

We measured the dependence of I2C bus power dissipation
on the pull-up resistance at $f_\mathrm{I2C}=200\,\mathrm{kHz}$.
We selected $3.92\,\mathrm{k}\Omega$, $7.84\,\mathrm{k}\Omega$,
$15.69\,\mathrm{k}\Omega$, $31.37\,\mathrm{k}\Omega$, and
$62.75\,\mathrm{k}\Omega$ as pull-up resistance values for
investigation. The lowest value is in the range of typical
external I2C pull-up resistance values.

We used an MMA8451Q digital
accelerometer~\cite{MMA8451Q_digital_accelerometer} as the target
sensor in our measurements. To control the bytes transmitted
by the sensor over the I2C bus, we write a prescribed value to a sensor 
configuration register and request the sensor to send that value back $60\,000$
times.  We used two sets of bytes to characterize the power dissipation
on the I2C bus. The first set (\code{0000\,0000}, \code{1100\,0000},
\code{1111\,0000}, \code{1111\,1100}, \code{1111\,1111}) has bytes with
varying duty cycles of logic~\logiczero and \logicone and have fixed logic switch
frequency. The second set of bytes (\code{1010\,1010}, \code{1100\,1100},
\code{1111\,0000}) all have equal duty cycles for both logic values but
have varying number of transitions between logic states.

We use a Keithley SMU2450 6.5-digit laboratory source-measure unit with a current
measurement resolution of 10\,nA to power the Warp and determine
the power consumed by I2C communication by measuring
the current drawn by the power supply. Figure~\ref{fig:I2C-experiment}(a)
shows raw current measurement data at the lowest and highest DCP
settings. Each data point corresponds to a current measurement.
The first five high current plateaus correspond (in order) to
transmissions of the five bytes in the first set given above and the last
two correspond (in order) to the first two bytes in the second set.

From the raw current data we estimate average current consumption
during repeated I2C transmissions of a given byte by taking the
difference between the averages of data points on the plateau and
the averages of points around it (four on each side). We then find
the average power consumption by multiplying the average current
consumed by the positive supply voltage, which we set to
$2.4\,\mathrm{V}$ in the experiment.

\subsection{Measurements Results}
\label{subsection:I2C-measurement-results}
Figure~\ref{fig:I2C-experiment}(b)~and~(c) show the dependence
of I2C power consumption on the I2C pull-up resistance value for
transmitted bytes with varying duty cycles (and fixed logic switch
frequency) and with varying logic switch frequency (and fixed duty
cycle), respectively. The results show a clear inverse dependence
of the I2C power dissipation on the pull-up resistance value with
an absolute value of power savings of up to $0.8\,\mathrm{mW}$. The
power dissipation, which can be as high as $2\,\mathrm{mW}$ for the
lowest used resistance value of $3.92\,\mathrm{k}\Omega$ (typically used in practice),
reduces below $1.2\,\mathrm{mW}$ and settles to a plateau above
$1\,\mathrm{mW}$, implying up to 2$\times$ power reduction.
The reason for this plateau is the existence of the internal pull-up
pathways embedded into sensors and/or processors as
Figure~\ref{fig:I2C-schematic} shows. Furthermore, the results confirm
the expected behavior of increasing power consumption with increasing
duty cycle of logic~\logiczero and reveal the increasing power
consumption with increasing logic switch frequency.

Because our experimental procedure involved a repetitive register
read request of a fixed byte from a sensor register, the actual
reading process was only a portion of the total I2C transaction.
As a result, the observed differences in current measurements only
reflect changes in power dissipation due to the transmission of a prefixed
byte during that portion of the whole process. Consequently, observed
power consumption dependencies in Figure~\ref{fig:I2C-experiment}(b)~and~(c)
are less than what would be in a real-word usage scenario, where a sensor
transmits streams of uninterrupted data.

\subsection{Estimation of I2C High-Logic-Reception Channel Model Parameters}
\label{subsection:I2C-channel-estimation}
Both the analysis of I2C circuitry (shown in Figure~\ref{fig:I2C-schematic}(a))
and the experimental results indicate that one can decrease the
power consumption of I2C communication by increasing the pull-up
resistance of the I2C data bus. Similarly, it follows that an increase
in pull-up resistance will only have detrimental effects on the
transmission fidelity of high logic values. Specifically, it results in a
decrease of the steady-state logic~\logicone I2C voltage level
(bringing it closer to threshold) and in an increase of \logiczero-to-\logicone
transition time (risking reaching to steady-state level in an I2C clock cycle).
In stark contrast, an increase in pull-up resistance causes a
decrease of the steady-state logic~\logiczero I2C voltage level
(taking it further away from threshold) and a decrease of
\logicone-to-\logiczero transition time. Based on this observation,
which we also verified with measurements provided below, the potential
of our proposed method lies in the analysis of logic~\logicone transmission.

We evaluated the two extreme
configurations of the pull-up DCP by measuring the steady-state
logic~\logiczero/\logicone voltage levels $V^{\logiczero/\logicone,\mathrm{min}}_\mathrm{I2C}$
and $V^{\logiczero/\logicone,\mathrm{max}}_\mathrm{I2C}$,
as well as the rise/fall times ($10\%$--$90\%$) of the I2C voltage,
for the minimum ($R_{\mathrm{I2C}}^{\mathrm{pu},min}\!\!\approx \!3.9\,\mathrm{k}\Omega$)
and maximum ($R_{\mathrm{I2C}}^{\mathrm{pu},max}\!\!\approx \!62.7\,\mathrm{k}\Omega$)
DCP pull-up resistance configurations, respectively. 

\begin{figure}[t]
\centering
{\includegraphics[trim=1cm 0.3cm 1cm 0cm, clip=true, angle=0, width=0.5\textwidth]{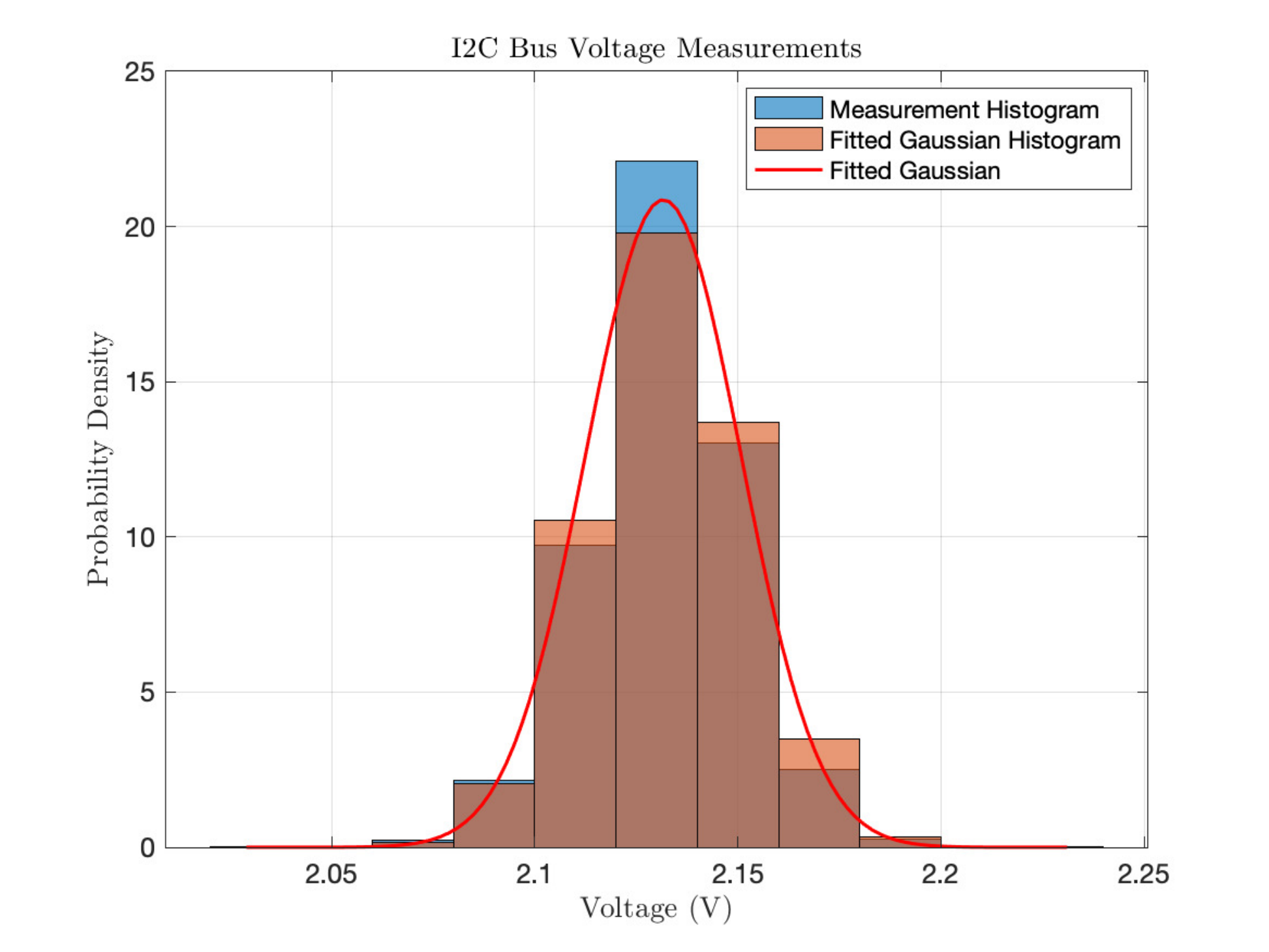}}
\caption{Density histogram (the area under the histogram integrates to unity) of voltage measurements of the I2C bus with the minimum DCP pull-up resistance configuration ($44\,235$ samples), the fitted Gaussian distribution with same mean and standard deviation as the data and the density histogram of samples drawn from the fitted Gaussian with same resolution as the measurement data. Two-sample Cram\'{e}r-von Mises test rejects the null hypothesis that the measurement data is distributed as the samples from the fitted Gaussian at the $1\%$ level.}
\label{fig:i2cVoltageGaussianity}
\end{figure}

Figure~\ref{fig:i2cVoltageGaussianity} plots the distribution of the
steady state voltage measurements of the I2C bus with the minimum
DCP pull-up resistance configuration at supply voltage $V_\mathrm{supply}=2.5\,\mathrm{V}$,
together with the fitted Gaussian distribution with the same mean and standard deviation
as the measurement data and the density histogram of samples drawn from the fitted Gaussian.
Even though the plot visually confirms that the density histogram of the measured steady-state
voltage of the I2C bus is close to the density histogram of samples drawn from the fitted Gaussian,
the two-sample Cram\'{e}r-von Mises test rejects the null hypothesis that
the measurement data is distributed as the samples drawn from the fitted Gaussian
at the $1\%$ level. Measurements yielded the
following means and standard deviations for the low and high logic voltage levels
of the I2C bus with the minimum and maximum DCP pull-up resistance configurations:
\begin{equation*}
\begin{split}
&V^{\logiczero,\mathrm{min}}_\mathrm{I2C}=58.9\,\pm\,19.4\,\mathrm{mV},\quad
V^{\logicone,\mathrm{min}}_\mathrm{I2C}=2131.6\,\pm\,19.1\,\mathrm{mV},\\
&V^{\logiczero,\mathrm{max}}_\mathrm{I2C}=25.2\,\pm\,10.0\,\mathrm{mV},\quad
V^{\logicone,\mathrm{max}}_\mathrm{I2C}=2072.3\,\pm\,19.6\,\mathrm{mV},
\end{split}
\end{equation*}
and rise/fall times
\begin{equation*}
\begin{split}
&\tau_\mathrm{rise}^\mathrm{min}\approx 454\,\mathrm{ns},
\quad\quad\,\,\,\, \tau_\mathrm{fall}^\mathrm{min}\approx 18\,\mathrm{ns},\\
&\tau_\mathrm{rise}^\mathrm{max}\approx 1092\,\mathrm{ns},
\quad\quad \tau_\mathrm{fall}^\mathrm{max}\approx 17\,\mathrm{ns}.
\end{split}
\end{equation*}
These results verify the previous hypothesis that increasing pull-up resistance
increases the rise time and decreases the fall time of the I2C bus voltage.

We denote the equivalent resistance of the pull-up branch in Figure~\ref{fig:I2C-schematic}(b)
when the DCP is in minimum/maximum pull-up resistance configuration by
$$
R_\mathrm{eq}^{\mathrm{pu},\mathrm{min/max}}=R_\mathrm{eq}^\mathrm{ipu}\parallel R_{\mathrm{I2C}}^{\mathrm{pu},\mathrm{min/max}}.
$$
$R_\mathrm{eq}^\mathrm{ipu}$ is the equivalent resistance for the internal pull-up
paths (connected to the supply in parallel) of all devices on the I2C bus.
The resistance $R_\mathrm{eq}^\mathrm{off}$ in the pull-down path of the equivalent circuit in
Figure~\ref{fig:I2C-schematic}(b) is the equivalent resistance for the off-resistances of the internal
transistors in the pull-down path (connected to the ground in parallel) of all devices on I2C bus.

The steady-state voltage level measurements provided above let us estimate all the resistances
in Figure~\ref{fig:I2C-schematic}(b). In particular,
\begin{equation}\label{eqn:measurement-ratios}
\frac{R_\mathrm{eq}^{\mathrm{pu},\mathrm{min/max}}}{R_\mathrm{eq}^\mathrm{off}}=
\frac{V_\mathrm{supply}-V^{\logicone,\mathrm{min/max}}_\mathrm{I2C}}{V^{\logicone,\mathrm{min/max}}_\mathrm{I2C}}.
\end{equation}
There are two equalities in~\eqref{eqn:measurement-ratios}. Taking the ratio of these equalities
and plugging in the steady state voltage values gives
$$
\frac{R_\mathrm{eq}^\mathrm{ipu}\parallel R_{\mathrm{I2C}}^{\mathrm{pu},\mathrm{max}}}{R_\mathrm{eq}^\mathrm{ipu}\parallel R_{\mathrm{I2C}}^{\mathrm{pu},\mathrm{min}}}
=\frac{R_\mathrm{eq}^{\mathrm{pu},\mathrm{max}}}{R_\mathrm{eq}^{\mathrm{pu},\mathrm{min}}}\approx 1.19.
$$
With the knowledge of the DCP pull-up resistance values $R_{\mathrm{I2C}}^{\mathrm{pu},\mathrm{min/max}}$
this ratio yields an equivalent internal pull-up resistance of $R_\mathrm{eq}^\mathrm{ipu}\approx \!0.82\,\mathrm{k}\Omega$,
which in turn yields an equivalent transistor off-resistance of $R_\mathrm{eq}^\mathrm{off}\approx \!3.94\,\mathrm{k}\Omega$.

After estimating the resistance values we can estimate $C_\mathrm{I2C}$ using the rise/fall time measurements
and the analytic solution of the voltage on the I2C data bus,
\begin{equation}\label{eqn:I2C-voltage-analytic-solution}
\begin{split}
V_\mathrm{I2C}(t)=&\,\,V_\mathrm{I2C}(0)e^{-\frac{t}{\tau_\parallel^\mathrm{min/max}}} \\
&+V^{\logicone,\mathrm{min/max}}_\mathrm{I2C}
\left(1-e^{-\frac{t}{\tau_\parallel^\mathrm{min/max}}} \right),
\end{split}
\end{equation}
where $\tau_\parallel^\mathrm{min/max}=R_\parallel^\mathrm{min/max} C_\mathrm{I2C}$ and
$R_\parallel^\mathrm{min/max} = R_\mathrm{eq}^{\mathrm{off}}\parallel R_\mathrm{eq}^{\mathrm{pu},\mathrm{min/max}}$.
Using this equation we estimate the I2C capacitance of our system as $C_\mathrm{I2C}^\mathrm{min}\approx 522\,\mathrm{pF}$.

\subsection{Analytic Derivation of Dependence of I2C Channel Integer Value Distortion on Pull-up Resistance}\label{subsection:I2C-IVD-analytic-derivation}

\begin{figure*}
\centering
\subfloat[Induced I2C errors are sensitive to $R_\mathrm{eq}^\mathrm{pu}$. Consecutive settings of the DCP in \Warp ($8$--$12$) are too coarse to have control over error dynamics. The straight line is the worst possible operation, where all \logicone's are wrongly perceived as \logiczero.]{\includegraphics[trim=0cm 0cm 0cm 0cm, clip=true, angle=0, width=0.33\textwidth]{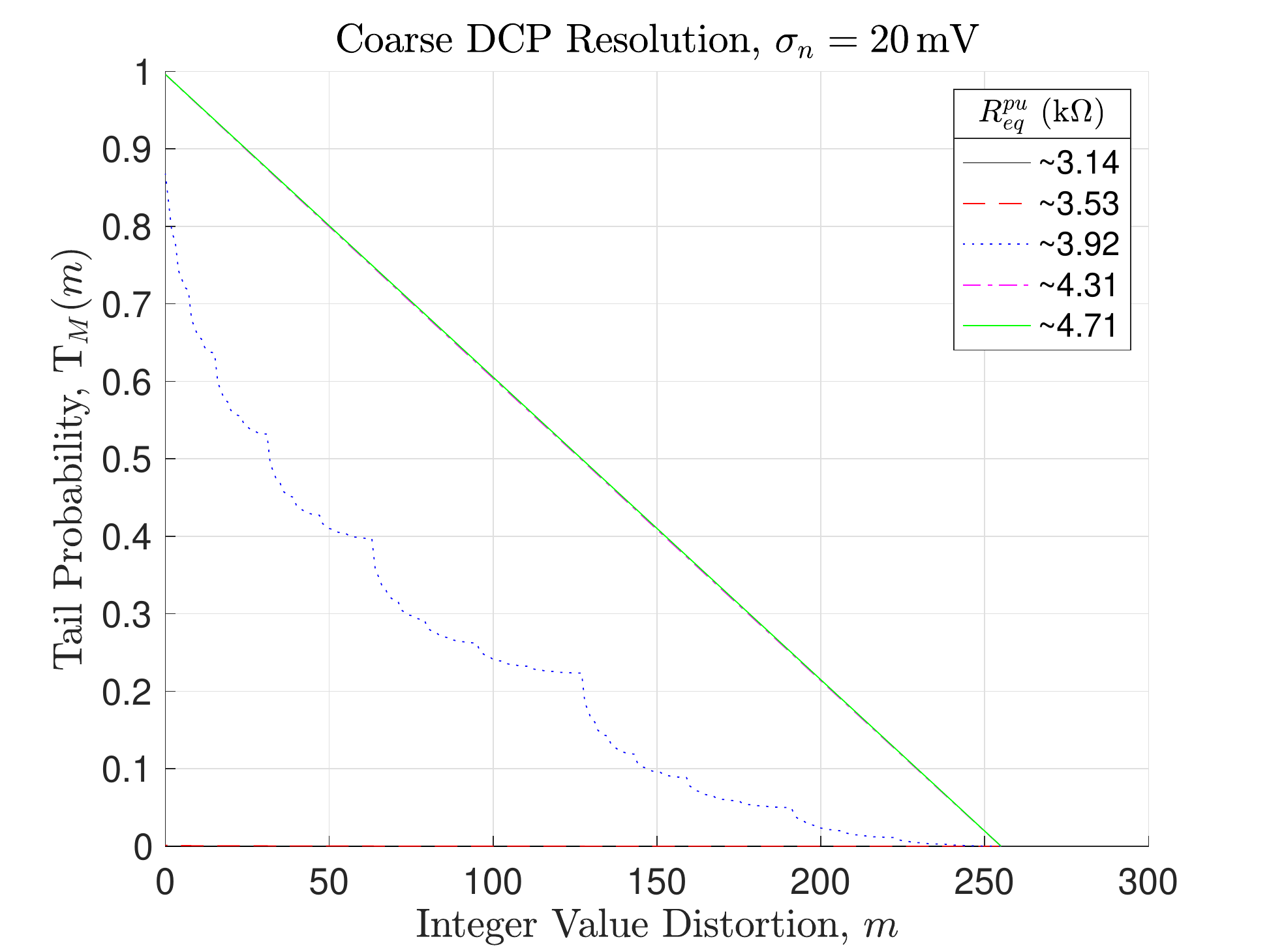}}~~
\subfloat[Increasing the resolution of the DCP ($10\times$) improves control over induced I2C error characteristics. The resistance values vary around the same value ($3.92\,\mathrm{k}\Omega$) as in (a).]{\includegraphics[trim=0cm 0cm 0cm 0cm, clip=true, angle=0, width=0.33\textwidth]{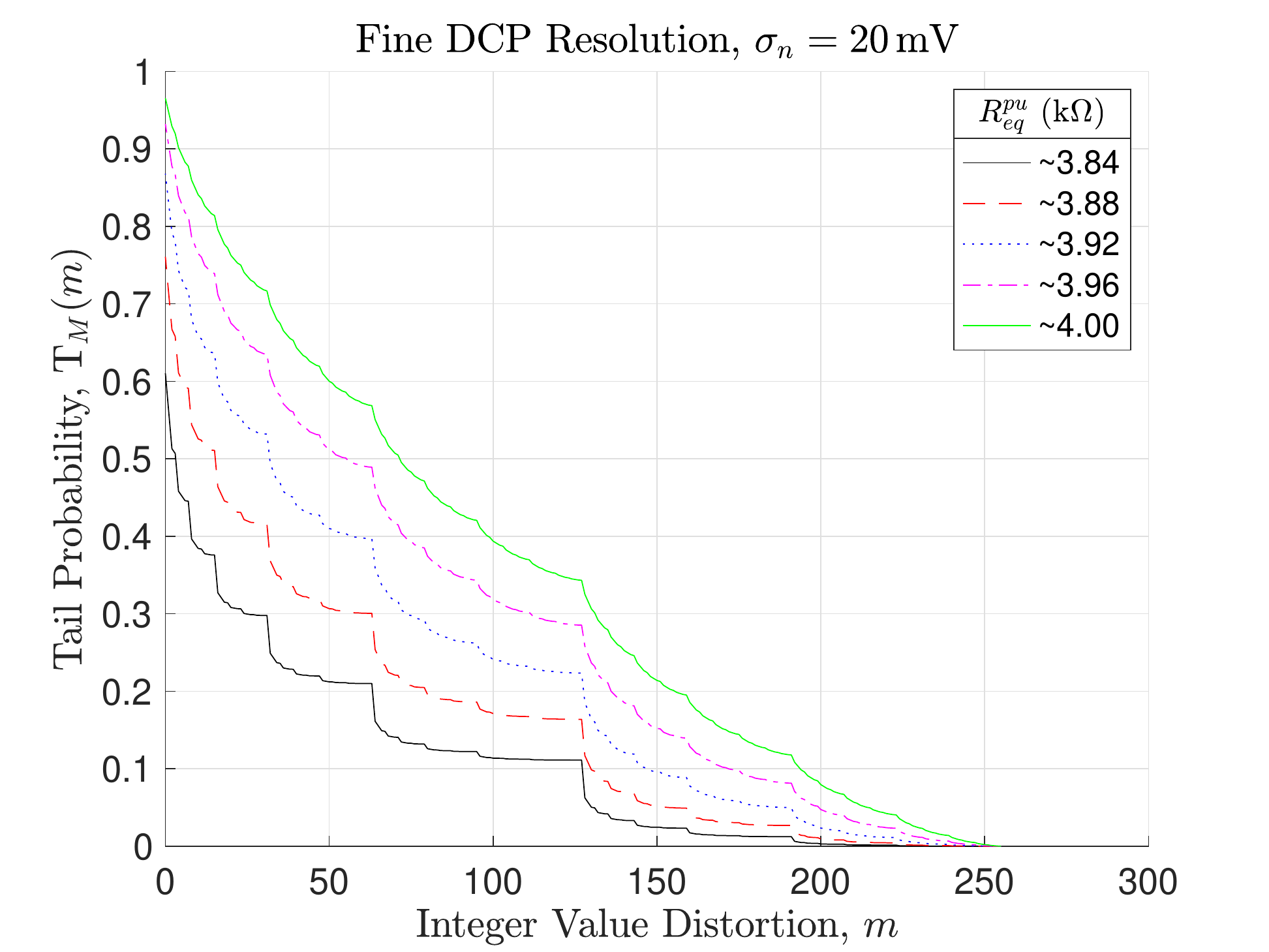}}~~
\subfloat[Induced I2C errors are sensitive to the noise levels on the I2C channel. Thus, control over I2C errors induced requires monitoring of and compensation for noise levels.]{\includegraphics[trim=0cm 0cm 0cm 0cm, clip=true, angle=0, width=0.33\textwidth]{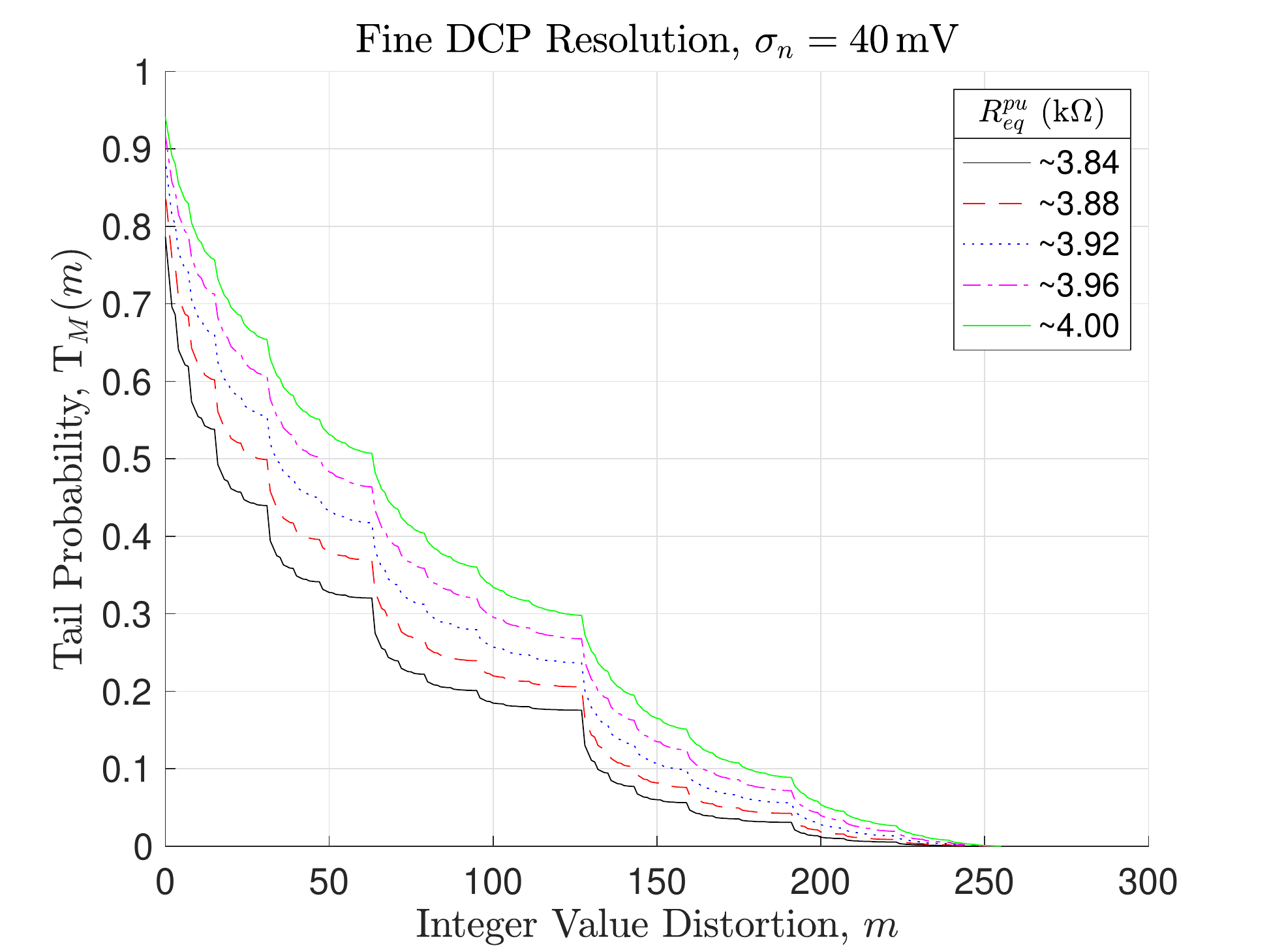}}
\caption{Analysis of tail distributions of integer value distortion suffered in I2C channels for uniform input distribution. I2C error dynamics are highly sensitive to the equivalent pull-up resistance value $R_\mathrm{eq}^\mathrm{pu}$ and the standard deviation $\sigma_n$ of the noise on I2C channel.}
\label{fig:I2C-error-analysis}
\end{figure*}

During bit reception process over I2C from a sensor, the sensor
sets the voltage of the I2C bus at the beginning of an I2C clock cycle
(when the I2C clock SCL is low), either pulling the SDA voltage down
to logic~\logiczero or letting the supply pull the SDA voltage up. We assume
that an erroneous bit read from a sensor on I2C occurs when the I2C
voltage sampled by the microprocessor at the end of the clock cycle
(during the high-to-low transition of SCL) is on the wrong side of the
threshold voltage $V_\mathrm{th}$. We take it to be half of the supply
voltage, i.e., $V_\mathrm{th}=\frac{V_\mathrm{supply}}{2}$. The possible
causes for such an event can be long transition times between logic
levels compared to the I2C clock period $t_\mathrm{I2C}$ and logic levels
which are on the wrong side of $V_\mathrm{th}$, possibly due to noise.

Let $R_\mathrm{eq}^{\mathrm{pu},i}$ denote the equivalent
resistance on the total pull-up path in the equivalent circuit in
Figure~\ref{fig:I2C-schematic}(b) when the DCP is at the $i$'th
setting. The ISL23415 DCP we use in the experiments of Section~\ref{subsection:I2C-measurement-results}
has 256 programmable levels, so $0\leq i \leq 255$.
Let $V^{\logiczero/\logicone,i}_\mathrm{I2C}$
be the corresponding steady-state logic~\logiczero/\logicone voltage
levels on the I2C bus and let $\tau_i$ be the corresponding rise time constant.
Let $\{i_1,\ldots ,i_8\}$ denote the DCP setting values for the pull-up
resistances we switch through during transmission of each integer-encoding byte,
where $i_1$ corresponds to the first bit of the byte sent over I2C.
This bit is the least or most significant bit of the byte if the sensor transmits
data in little- or big-endian fashion, respectively.
Also, let $0\leq i_\mathrm{nom}\leq 255$ be the nominal index
for the DCP setting with corresponding equivalent pull-up resistance
$R_\mathrm{eq}^{\mathrm{pu},i_\mathrm{nom}}$
yielding reliable operation without any read errors. We assume that the
DCP module switches to this resistance for operation-critical bus transactions,
e.g., during transmission of START/STOP or ACK/NACK, and bytes which
are not integer-encoded or which cannot tolerate non-zero integer value distortion.

During data reception on I2C from a sensor, the sensor sequentially
transmits sensor data in chunks of bytes. During the clock cycle, just
before each such byte transmission, SDA is always low. Either the sensor
sets it low just before the transmission of the initial byte as an acknowledgement
(ACK) to the read request from the microprocessor or the master sets it low
as an ACK in between two consecutive bytes the sensor transmits. Since
logic~\logiczero transmission on I2C is always reliable, at the beginning
of transmission of each integer-encoding byte the SDA voltage will be
steady at $V^{\logiczero,i_\mathrm{nom}}_\mathrm{I2C}$.

For $1\leq j\leq 8$ and a given byte $x$ sent over the I2C bus let $x(j)$ denote the
value of $j^{th}$ transmitted bit (in time). In the I2C protocol the most significant
bit is transmitted first, so that with the notation introduced in Section~\ref{section:definitions}
we have $x(1)=x_7$. We find the probabilities of erroneous bit reception
during transmission of $x$ given the DCP configuration
profile $\{i_j\}_{j=1}^8$ via the following iterative calculations.

Assuming that the transmission of $x$ occurs during the time interval
$[0,8*t_\mathrm{I2C}]$ and that the initial I2C voltage at time $t=0$ is
$V_\mathrm{I2C}(0)=V^{\logiczero,i_\mathrm{nom}}_\mathrm{I2C}$,
iteratively on $1\leq j\leq 8$ we solve an equation similar to the
Equation~\eqref{eqn:I2C-voltage-analytic-solution} over the time interval
$[(j-1)t_\mathrm{I2C},jt_\mathrm{I2C}]$ to find $V_\mathrm{I2C}(jt_\mathrm{I2C})$.
If $x(j)=\logiczero$, then transitions are rapid and I2C voltage reaches
steady-state level before the end of clock cycle, so that at the end of the
$j$'th cycle we set the I2C voltage to be $V_\mathrm{I2C}(jt_\mathrm{I2C})=V^{0,i_j}_\mathrm{I2C}$.
If $x(j)=\logicone$ however, then transitions are slow and we calculate the
voltage levels at the end of each clock cycle as
\begin{equation*}
\begin{split}
V_j:&=V_\mathrm{I2C}(jt_\mathrm{I2C})\\
&=V_\mathrm{I2C}((j-1)t_\mathrm{I2C})e^{-\frac{t_\mathrm{I2C}}{\tau_{i_j}}}
+V^{\logicone,i_j}_\mathrm{I2C}
\left(1-e^{-\frac{t_\mathrm{I2C}}{\tau_{i_j}}} \right).
\end{split}
\end{equation*}

We take the I2C voltage levels at the end of each I2C clock cycle
to be Gaussian distributed (see Figure~\ref{fig:i2cVoltageGaussianity})
with mean $V_j$ and standard deviation $\sigma_n$, $1\leq j\leq 8$.
Then, given $x(j)=\logicone$, the probability that the receiver erroneously
perceives it as logic~\logiczero is
$$
p^x_j=\frac{1}{\sigma_n\sqrt{2\pi}}\int\limits_{V\leq V_\mathrm{th}} e^{-\frac12\left(\frac{V-V_j}{\sigma_n}\right)^2}\,dV.
$$
After appropriate relabeling of indices (from $1\leq j\leq8$ to $0\leq i\leq7$)
the transmitted-byte-dependent error probabilities of Equation~\eqref{eqn:p_i}
in Section~\ref{subsection:calculation-IVD} read
$$
p^x_{i,\mathrm{down}}=p^x_j,\quad x_i=x(j);
$$
and
$$
p^x_{i,\mathrm{up}}=0, \quad \forall x\in\mathbb{W}_8,\quad 0\leq i\leq7.
$$
Finally, using these probabilities we calculate the distribution of the integer value distortion of the channel
with the given DCP configuration profile as described in Section~\ref{subsection:calculation-IVD}.

Figure~\ref{fig:I2C-error-analysis} shows the dependence of the tail
distribution of integer value distortion $\mathrm{T}_{\!M}$ on the equivalent
I2C pull-up resistance values $R^\mathrm{pu}_\mathrm{eq}$ in the case
of uniform input distribution. Each plot corresponds to a DCP configuration
profile with a fixed resistance across a whole byte. The fixed resistances
correspond to the DCP settings used in power measurement experiments
of Section~\ref{subsection:I2C-power-dissipation}. We choose the channel
parameters same as the values estimated for our system in Section~\ref{subsection:I2C-channel-estimation}
with the exception of the bus capacitance, which we set equal to
$C_\mathrm{I2C}=100\,\mathrm{pF}$, more representative of an embedded
system with fewer sensors connected to the I2C bus than in the Warp experimental
system described in Section~\ref{subsection:I2C-power-dissipation}.
We assume devices on I2C have no internal pull-up
paths, so that the external DCP pull-up dominates the pull-up resistance.

Figure~\ref{fig:I2C-error-analysis}(a) shows the distribution of induced
integer value distortion for the DCP (ISL23415TFUZ) on the Warp system used
in the power dissipation experiments described in Section~\ref{subsection:I2C-power-dissipation}.
The resolution ($0.39\,\mathrm{k}\Omega$) of the ISL23415TFUZ ($100\,\mathrm{k}\Omega$)
was too coarse to have the desired accuracy of I2C error manipulation.
Resistance values corresponding to consecutive DCP settings ($8$--$12$)
yielded big jumps in the error performance of the I2C channel.
However, Warp also supports the DCP ISL23415WFUZ ($10\,\mathrm{k}\Omega$),
which is pin compatible with the ISL23415TFUZ and has $10\times$ finer
resolution. Figures~\ref{fig:I2C-error-analysis}(b)~and~(c) demonstrate the
improved control on the I2C error one can achieve when using the ISL23415WFUZ
as the choice of the DCP. There is no fundamental reason for not using a DCP
with an even smaller granularity in a custom silicon implementation.

Figures~\ref{fig:I2C-error-analysis}(b)~and~(c) also
demonstrate the sensitivity of the behavior of I2C error to the I2C noise levels.
Figure~\ref{fig:I2C-error-analysis}(b) shows results of calculations for I2C
voltage noise standard deviation $\sigma_n=20\,\mathrm{mV}$, whereas
Figure~\ref{fig:I2C-error-analysis}(c) shows results for $\sigma_n=40\,\mathrm{mV}$.
As expected, increased noise deteriorates the control over the I2C error
and shifts the tail distribution of integer value distortion of the channel upwards.
Thus, control of the induced I2C channel errors requires one to monitor the
noise levels on I2C bus and compensate for changes in noise levels.
Because primarily \logicone~$\rightarrow$~\logiczero errors can occur as a
direct result of increasing pull-up resistor value,
I2C channel errors result in received values that are numerically smaller than the
actual value transmitted and the integer value distortion
(which by definition is nonnegative) shown in Figure~\ref{fig:I2C-error-analysis}
correspond to negative errors only.
\section{Hardware Overheads of Channel Adaptation}
\label{section:hardware}
In Section~\ref{section:I2C} we implemented channel adaptation
to I2C via a DCP module. In general, channel adaptation applies
to any binary channel that transmits integer data and will require
a channel adaptation module similar to the DCP module for each
such channel. In this section we estimate the hardware cost of the
channel adaptation module that is independent of the channel
communication protocol for serial communication channels such as
I2C and SPI, which are widely used in embedded systems.

\begin{figure}[t]
\centering
{\includegraphics[trim=1cm 3cm 1cm 3cm, clip=true, angle=0, width=0.5\textwidth]{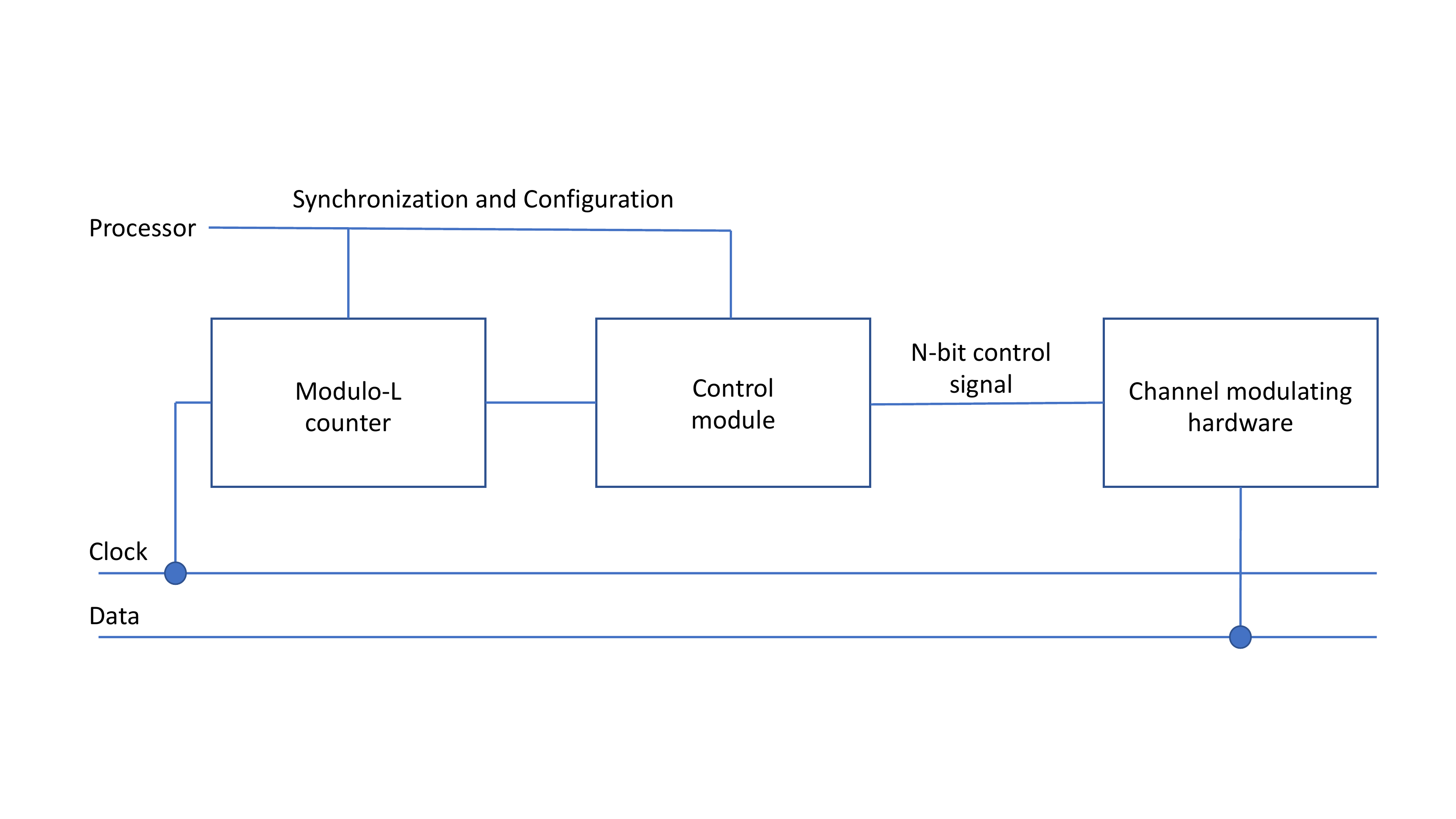}}
\caption{The channel adaptation module consists of a channel modulating hardware (e.g., a DCP) facilitating adaptation of the data bus, an $L$-bit counter and a control module that provides the $N$-bit control signal to the channel modulating hardware The processor feeds the configuration parameters to the control module only once and synchronizes the modulo-$L$ counter with the serial communication transactions.}
\label{fig:channel_adaptation_module}
\end{figure}

Utilization of bit-level channel adaptation exhibits a number
of technical complications from the processor side, such as
the operational load introduced to the processor, its accompanying
power consumption and the difficulty in implementation from the timing
perspective. However, these are straightforward to mitigate by providing a
channel adaptation module that acts as the glue logic between
the processor and the channel modulating hardware, e.g., the DCP module
in the case of I2C communication discussed in Section~\ref{section:I2C}.

Figure~\ref{fig:channel_adaptation_module} illustrates a block diagram
for the architecture of such a channel adaptation module.
For a single sensor transmitting $\wordLength$-bit words on a serial
channel,  $L+1$ $N$-bit registers (1 register for each bit position and
1 for default value) suffice to store a bit-level channel adaptation
configuration, which consists of selection values to choose from the
$2^N$ possible states the channel modulating hardware can assume.
The approach of feeding these values constantly through the
CPU is energy inefficient and does not meet the the timing requirements
of bit-level adaptation. To control the channel modulating hardware we
propose a minimal control module implemented in a miniature low-power FPGA,
which can be easily integrated, for instance, into existing DCP designs.
The module receives from the processor the channel adaptation lookup
table only once and executes adaptation when instructed. This enables
fast and low-power switching of the channel modulating hardware and
facilitates bit-level adaptation of high-frequency serial communication channels.

To quantify the hardware footprint of the protocol-agnostic part of the channel
adaptation module, we implement its control logic in an FPGA using the
Verilog hardware description language. Algorithm~\ref{alg:Channel-Adaptation-Module-Verilog-Implementation}
encapsulates our implementation. We design the hardware module as a
two-state finite state machine that modulates the communication channel
at each negative edge of the transmission clock during data transaction.
During the transmission of the $i$'th bit of a word, the module modulates
the channel according to the selection specified by register $R_i$.

We have synthesized the Verilog implementation on an iCE40 FPGA for
word lengths $\wordLength=8\,/12\,/16$. The synthesis revealed that, the
protocol-agnostic part of the synchronization logic for the channel adaptation
module utilizes $61\,/61\,/61$ carry logic, $34\,/34\,/34$ negative edge clock
D flip-flops, $72\,/104\,/136$ negative edge clock D flip-flops with clock enable
and $224\,/248\,/291$ $4$-input LUTs.

\begin{algorithm}
\footnotesize
\DontPrintSemicolon
\SetKwInput{KwInput}{Inputs}
\SetKwInput{KwOutput}{Outputs}
\SetKwInput{KwHardware}{Hardware Components}
\SetKwInput{KwStates}{State Transitions}
\SetKwInput{KwAlways}{Always}

\KwInput{$SCL$ (clock signal), $S_\mathrm{start}$ (start signal) and $\wordLength$ (word length).}

\KwOutput{$S_\mathrm{select}$ (an $8$-bit selection signal to manipulate channel modulation hardware.}

\KwHardware{
\par
-- A channel modulating hardware with $8$-bit selection option.

-- $L+1$ $8$-bit registers ($R_0$ to $R_L$) to hold the $L+1$ channel modulation selections.
$R_0$ holds the default selection value and $R_1$ to $R_L$ hold the selections to switch through
during sensor data transmission.
}

\KwStates{
\par
-- Start at the initial state $s_0$\\
-- \textbf{In $s_0$:} \tcp*{The idle state}\\
i=0;\\
\tcc{Beginning of a word transmission}
\If{$S_\mathrm{start}==1$ and $SCL$ at negative edge}
{
i=i+1\\
Switch to $s_1$\\
}
\noindent
-- \textbf{In $s_1$:} \tcp*{Counting state}\\
\If{$i<L$ and $SCL$ at negative edge}
{
i=i+1\\
\ElseIf{$i=L$ and $SCL$ at negative edge}
{
i=0\\
Switch to $s_0$\\
}
}
}

\KwAlways{
$S_\mathrm{select}\leftarrow R_i$\\
}

\normalsize
\caption{Pseudocode for synchronizing the channel adaptation module to a serial channel with transmitted words of size $\wordLength$.}
\label{alg:Channel-Adaptation-Module-Verilog-Implementation}
\end{algorithm}
\section{Conclusions}
\label{section:conclusions}
We present a new communication channel adaptation technique
that trades integer distance in induced errors for lower power usage.
The technique achieves this by modulating the channel to vary the
bit-error rate across the ordinal bit positions in a word to reduce
power dissipation. To predict channel parameters, it utilizes an efficient
method for calculating the exact distribution of integer value distortion
suffered during bit-level adaptation. We carry out a case study for
implementation of the technique on I2C channel. We empirically
verify power reduction of up to 2$\times$ ($0.8\,\mathrm{mW}$) by
means of modulating the pull-up resistance of the I2C bus of the
\Warp hardware platform. We explore the design space and feasibility
of the proposed scheme utilizing the developed theoretical tools for
calculating the integer value distortion distribution and analytical derivations
of I2C channel errors. Results indicate the need for noise compensation
for reliable channel adaptation. Finally, by means of synthesizing
in FPGA the protocol-agnostic part of bit-level synchronization to serial
communication channels, we give an estimate for the footprint of a channel
modulation hardware for sensor sample resolutions common in embedded
sensor systems (e.g., $8$--$16$ bits).

\vspace{-0.1in}
\ifCLASSOPTIONcompsoc
  \section*{Acknowledgments}
\else
  \section*{Acknowledgment}
\fi

This research is supported by an Alan Turing Institute award
TU/B/000096 under EPSRC grant EP/N510129/1, and by Royal Society
grant RG170136. We thank Richard Hopper for
assisting with power dissipation measurements in Section~\ref{section:I2C},
Srisht Fateh Singh for help with the FPGA synthesis in
Section~\ref{section:hardware}, and James Meech, Nathaniel Tye,
Chatura Samarakoon, and Vasileios Tsoutsouras for providing feedback
on the manuscript.

\ifCLASSOPTIONcaptionsoff
  \newpage
\fi

\bibliographystyle{IEEEtran}
\bibliography{probabilistic-vdbe}
\end{document}